\documentclass[12pt]{article}
\usepackage{graphicx}
\usepackage{tabularx, booktabs}
\newcolumntype{Y}{>{\centering\arraybackslash}X}
\newcolumntype{B}{>{\centering\arraybackslash\hsize=1.5\hsize}X}
\newcolumntype{Q}{>{\centering\arraybackslash\hsize=2\hsize}X}
\usepackage{titlesec}
\usepackage{wrapfig}
\usepackage{epstopdf}
\usepackage{setspace}
\usepackage[left=1in,top=1in,right=1in,bottom=1in,nohead,paperwidth=8.5in, paperheight=11in]{geometry} 

\usepackage{amsmath}
\usepackage[title,titletoc,toc]{appendix}
\usepackage{color}
\usepackage{caption, subcaption}

\usepackage[round]{natbib}
\usepackage{lipsum}
\usepackage{indentfirst}
\usepackage{hyperref}          
\hypersetup{
	colorlinks=true,
	citecolor=blue
}
\graphicspath{ {Figures/} }

\usepackage{dsfont}

\usepackage{algorithm}
\usepackage{algorithmic}
\usepackage{leading}

\usepackage{setspace}

\title{\vspace{-1.0cm}  \bf Monitoring Machine Learning Forecasts for Platform Data Streams\footnote{Correspondence to 
Ines Wilms, Maastricht University, i.wilms@maastrichtuniversity.nl.
We are very grateful to Benjamin Wolter, Pablo Perez Piskunow and Roger Caminal for expert advice, to Aurore Delaigle, Geert Dhaene, David S.\ Matteson and Marie Ternes for comments and checks provided on earlier versions of the paper and to the participants at the Workshop ``Recent Advances in Econometrics" at UCLouvain,  the Data Science, Statistics \& Visualisation conference (DSSV - ECDA 2023), the IMS International Conference on Statistics and Data Science (ICSDS 2023), and the IMS Asia-Pacific Rim Meeting (APRM 2024)  for helpful discussions. IW was financially supported by the Dutch Research Council (NWO) under grant number VI.Vidi.211.032.}}

\author{
  Jeroen Rombouts$^a$ and  Ines Wilms$^b$ 
  \\ \textit{\small $^{a}$ Essec Business School, France}
  \\ \textit{\small $^{b}$ Department of Quantitative Economics, Maastricht University, The Netherlands}
}

\begin{document}
	
\begin{titlepage}
		\clearpage\thispagestyle{empty}
		\maketitle

		\begin{singlespace} 
  \noindent
  {\bf Abstract.} Data stream forecasts are essential inputs for 
  decision making 
  at digital platforms. Machine learning (ML) algorithms are appealing candidates to produce such forecasts. Yet,
  digital platforms require a large-scale forecast framework that can flexibly respond to sudden performance drops. Re-training ML algorithms at the same speed as new data batches enter is usually computationally too costly. On the other hand, infrequent re-training  requires specifying the re-training frequency 
  and typically comes with a severe cost of forecast deterioration.
To ensure accurate and stable forecasts, we propose a simple data-driven monitoring procedure to answer the question when the ML algorithm should be re-trained. Instead of investigating instability of the data streams, we test if the incoming streaming forecast loss batch 
differs from a well-defined reference
batch. 
Using a novel dataset constituting 15-min frequency data streams  from an on-demand logistics platform operating in London, we apply the monitoring procedure to popular ML algorithms including random forest, XGBoost and lasso. We show 
that  
monitor-based re-training produces accurate forecasts compared to viable benchmarks while preserving computational feasibility. Moreover, the choice of monitoring procedure 
is more important than the choice of ML algorithm, thereby permitting practitioners to combine the proposed monitoring procedure with one’s favorite forecasting algorithm.

\end{singlespace}

	\bigskip

			\noindent {\bf Keywords}: E-commerce; Platform econometrics; Machine Learning, Streaming data; Monitoring Forecasts.

		\thispagestyle{empty}
\end{titlepage}

\doublespacing

\clearpage

\section{Introduction} \label{Introduction}
The on-demand economy, created by digital market places fulfilling consumer demand via immediate access to goods and services, is rapidly expanding.
Digital platforms for on-demand services have contributed a sizable share to this growth with the gig economy invading multiple industries ranging from, amongst others,
ride-hailing (e.g., Uber, Lyft), 
over food deliveries (e.g., GrubHub, DoorDash), retailing (e.g., Instacart) to 
home services (e.g., TaskRabbit), and health services (e.g., Zeel). 
Currently, the gig economy is expanding much faster than the economy as a whole and
it hosts a significant share of the working population (for instance more than a third of the US workers, see \citealp{zgola21, wilson2023}).
 A key success factor for operating two-sided platforms concerns their ability to accurately forecast demand and supply, indispensable inputs when optimizing work planning and compensation schemes. However, platforms typically face irregular growth patterns which make forecasting performance unstable.
 In this paper, we develop a procedure for monitoring machine learning methods to ensure  accurate forecasts at 15-min intervals that can rapidly adapt to changing environments while preserving computational feasibility. We validate the procedure on a unique data set from a major on-demand logistics platform operating in London.

Forecast algorithms only create value for on-demand platforms when they are in production and used for decision making in real-time. Such forecasting is, however, complicated because of various statistical challenges.
First, platform data are generated continuously, typically with new batches entering at high velocity-- 15-min intervals in our application --and high granularity--  platforms operate in hundreds or even thousands of verticals  such as delivery areas, product categories, etc. --thus requiring forecasting at scale and subject to updating when new data streams arrive. Data with these characteristics are referred to as \textit{streaming data} in this paper. 
Second, \textit{speed} is vital at any on-demand digital platform.
Services are ``on-demand" meaning that upon experiencing a need for a service, customers desire the service immediately and are sensitive to delay (e.g., \citealp{taylor2018demand}) which causes a negative impact on the platform's  rating, a key performance indicator. 
For delivery platforms in particular, drivers-- who themselves decide to work or not --need to be online at locations with demand for deliveries. This puts pressure on the computational speed of the forecast algorithms since accurate, high-frequency forecasts are required to steer drivers at the right time towards the right locations,  and to determine the appropriate remuneration through dynamic pricing to balance supply (the drivers) and demand for deliveries (e.g., \citealp{garg2022driver}).
Third,  platform data streams are typically subject to frequent \textit{changes} since they highly depend on volatile market conditions due to competition, events and changing customer preferences (e.g., \citealp{irie2022sequential,hu2023fast}).
Forecast algorithms therefore need to rapidly adapt to prevent performance drops while accounting for computational constraints and ensuring (quasi-) automatic deployability at large scale.

Machine learning (ML) methods, with proven track record as winning contestants across various forecasting competitions (e.g., \citealp{makridakis2020m4, bojer2021kaggle, makridakis2022m5}) and successful economic applications (e.g., \citealp{Medeiros_2021, goulet2022machine, almeida2022can, masini2023machine}), are attractive to deliver forecasts in this challenging environment. 
They require little user intervention (up to tuning parameters), and sophisticated off-the-shelf implementations are routinely available thereby facilitating their operationalization in production environments. 
Furthermore, there is no limit on the number of features for a given target, feature selection and interaction are in-built, and for non-parametric ML methods one does not have to worry about the functional form of the forecast function. 
The downside is that ML algorithms can take considerable time to explore the space of possible forecast functions in order to minimize forecast errors. In addition, once a ML model is trained, the risk is non-negligible that new data is fundamentally different from the training data, therefore possibly producing large forecast errors if the model is not sequentially re-trained. In this light, producing large-scale, accurate yet timely forecasts with ML methods at on-demand platforms becomes a daunting task.

As training ML models is computationally costly in terms of time and resources, and re-training each time new data batches arrive is oftentimes simply impossible, this paper addresses the following question: When to re-train existing ML models to avoid systematic  deterioration of  forecast accuracy?
We propose a monitoring procedure, based on a standard statistical test, that triggers re-training when forecast performance-- as measured by a streaming loss function --changes compared to a well chosen reference batch.
Inspired by \cite{Luo_JASA_2022_batches}, we test for each demand data stream  whether its streaming forecast loss is equal, on average, to a reference batch consisting of the most recent historical period with stable forecast performance. Since our forecast loss streams are never in a baseline regime, we adapt their methodology by defining a new reference regime at each date a forecast performance shift is detected. 
Re-training is thus done at a priori unknown times and only when deemed necessary to avoid forecast deterioration due to a changing environment, thereby ensuring a good balance between computing time and forecast accuracy.

Thanks to the existing literature (Section \ref{Literature}), we have a variety of statistical tests  at our disposal with theoretically guaranteed correct size, and applicability to different types of data streams.
Our paper, however, differs from this literature because it addresses monitoring of \textit{forecast loss streams} generated by ML models. 
In fact, when forecasting demand at delivery platforms, the instability in the underlying dynamics of the data streams is of second order importance. What matters is the  forecast performance stability of the ML model.
The purpose of our monitoring procedure is to keep forecast accuracy high and re-train the algorithm only at times when the average forecast performance changes. Since forecast loss streams are highly nonlinear functions of the underlying data streams, dependent, and their inherent nonstationarity yields frequent forecast performance shifts, we extend the approach of \cite{Luo_JASA_2022_batches}, originally aimed at monitoring abnormal data batches. 

We apply our monitoring approach to a unique data set of an on-demand delivery platform facing 15-min demand (January 2019 to March 2021) and validate it across  the 32  areas that constitute its London market. 
As ML models, we consider parametric, regression-based ones with the lasso as popular representative and non-parametric ones with random forest and extreme gradient boosting (XGBoost) as representatives.
To make a fair comparison, each algorithm uses identical features consisting of trend, seasonality, and historical  demand for its own and all other delivery areas (to gauge the importance of spillovers). 
In terms of forecast accuracy, our monitoring scheme yields more accurate forecasts compared to monitoring based on a standard changepoint detection test, and this regardless of the ML algorithm used.
We compare the distributions of durations between re-training days for the three ML algorithms and find that the good performance of the proposed monitoring procedure comes from its tendency to re-train more frequently than the changepoint detection-based monitoring benchmark.
In terms of speed, monitoring-based re-training substantially reduces computing time compared to a deterministic daily re-training scheme, which would be too expensive to implement in a production environment.  
Finally, though we demonstrate the applicability of our monitoring procedure for on-demand platform forecasting, it can be used in a wide range of other applications that necessitate  monitoring of streaming forecasts subject to change; we provide an overview of such opportunities at the end of the paper.

The rest of the paper is organized as follows.
Section \ref{Literature} positions our contribution in the literature. 
Section \ref{PlatformData} introduces the platform data we analyze in this article.
Section \ref{Methodology} presents the forecast problem, our new procedure to monitor forecast performance and reviews the ML algorithms we use in combination with the monitoring procedure.
Section \ref{Forecasting} describes the forecast setup and benchmark procedures used in the application. 
Section \ref{sec:forecast_results} presents forecast performance results, monitoring test insights and discusses which parts of the ML model generate predictability.
Finally, Section \ref{Conclusion} concludes the article, suggests future research directions and discusses general applicability for other data streaming applications.

\section{Literature Review} \label{Literature}
Our paper is related to various disciplines; we briefly review related literature in economics, statistics and machine learning. 

In economics, the study of instabilities in time series, called structural breaks in econometrics, has a long tradition since \cite{Andrews_1993} and \cite{Bai_Perron_1998}, and 
more recently with \cite{PPT_RES_2006} proposing
models that incorporate future parameter breaks. Focusing directly on forecast stability,  \cite{Giacomini_Rossi_2009} have pioneered the ``forecast breakdown" research by proposing a forecast deterioration test that requires the entire historical time series; see \cite{Perron_JBES_2021} for recent improvements. However, such retrospective methods are not suitable in our continuously evolving streaming data setting in which only part of the time series is observable. Furthermore, a test result that points to a structural break does not indicate  what to do next. We refer to \cite{Rossi_JEL_2021} for an extensive econometrics literature survey on forecasting in the presence of instabilities.

In statistics, there is a vast literature on sequentially testing the compatibility  of an estimated  model when new data becomes available. 
\cite{PAGE_Biometrika_1955} is among the first to introduce a test for a change in a distribution parameter at unknown date based on cumulative sums (CUSUM) that are checked to exceed a threshold. 
\cite{White_ECA_1996} provides a real-time monitoring test for linear regression models based on CUSUM and fluctuation functions. Economic applications of this work and comparison with other procedures can be found in \cite{Zeileis_2005}, extensions for multivariate data streams in e.g., \cite{Mei_2010}, \cite{Dette_JASA_2020}. 
Besides, because instability can be caused by a misspecified parametric model, several nonparametric testing procedures have been proposed, see e.g.,  \cite{Chen_AOS_2019} and \cite{Padilla_JASA_2019}. 
Efficient online setting extensions and applications of CUSUM, for example,  to computer server data are available in e.g., \cite{Fearnhead_JMLR_2023} and \cite{Ward_JASA_2023}, whereas
\cite{qiu2020big} provides a recent review in the quality control domain.

Another important  closely related line of statistical research consists of monitoring data streams by sequentially testing a null hypothesis of absence of anomalies. \cite{Ross_2011_Technometrics}  adapts several nonparametric hypothesis tests to detect  changes in location or scale of streams of random variables and applies it to a financial data stream. 
  \cite{Gang_2023} provides anomaly detection methodology with a class of structure–adaptive sequential testing rules for online false discovery rate control with applications to time series and novel genotype identification. Note that in  this sequential testing setting, the null distribution can be estimated but is fixed and anomalies are detected with respect to this distribution throughout the  entire data stream; in contrast, in our setting the ``null distribution" (aka reference batch) frequently changes (see Section \ref{Methodology}). \cite{AUSTIN_2023_CSDA} also allows the null distribution to change over time, but thanks to sliding windows in their  nonparametric online changepoint detection method with an illustration to time series of telecom network devises.

Finally, in the ML literature, the topic of data shift and concept drift is getting increasingly  more attention, see e.g., \cite{suarez2022survey} for a survey. In fact, theoretical guarantees of ML algorithms are often derived for independently and identically (iid) sampled data, an assumption rarely met in practice. 
\cite{AHMAD2017134_Neurocomputing} propose a practical unsupervised real-time anomaly detection method for streaming data, but also theoretical frameworks emerge that study how to deal with shifts in the data distribution, see e.g.,  \cite{Federici_neurips_2021} and \cite{Zhang_neurips_2021}. In companies, data shift is part of MLops data-science tasks. MLops is a set of practices, bringing software development, ML and data engineering tools together to deploy and maintain ML systems in production. The MLops team monitors ML performance and acts when new incoming data differ from the existing data flow. Methodology wise, we refer to \cite{Rabanser_neurips_2019} for a survey on standard shift detection techniques. 

Our paper differs from these research lines because our monitoring scheme is aimed to detect forecast deterioration, i.e.\ a shift in forecast performance rather than data shifts. The focus on forecast performance simplifies the monitoring procedure (one forecast loss stream versus multiple data streams) and makes it easy to understand (bad forecasts can have severe business implications). Given that many companies are training staff of entire business units to become ``data citizens", it is crucial for adoption that a monitoring procedure is easy to explain, implement, and fast to run; our proposal offers these appealing features.

\section{On-Demand Platform Data} \label{PlatformData}
Our empirical analysis involves a unique data set of high-frequency demand data from a leading on-demand logistics platform in Europe, Stuart, which connects businesses to a fleet of geologically independent couriers.\footnote{The data are provided to us in the context of ongoing research collaborations. Due to a confidentiality agreement, we are not allowed to distribute or report actual demand data. Demand data across all figures are therefore normalized between 0-100. All analyses were carried out with the original data.} 
We analyze UK London demand data  at the 15-minute frequency from January 1, 2019 
to March 31, 2021.  London is split into 32  boroughs to efficiently organize parcel deliveries, the boroughs form a representative sample of the hundreds of delivery areas in its portfolio.\footnote{London is split into 32 boroughs plus City of London as administrative area. Since we have no data available for Hammersmith \& Fulham, for ease of exposition, we count 32 boroughs including City of London.}
To ease comparisons across boroughs and to allow for multivariate modeling, we assume that business operates everywhere on a daily basis between 9am and 11pm.\footnote{Occasional overnight demand is integrated in the last 15 minutes of the business day.} This creates a balanced dataset of 
$49,320$ 
observations per borough.
Fifteen-minute demand is aggregated over all product categories (food, fashion, health, etc.) because the proportions in these categories are relatively stable over our sample period. Borough demand forecasts  can therefore be easily decomposed into product  categories if necessary.

\begin{figure}[ht!]
\includegraphics[width=0.5\textwidth]{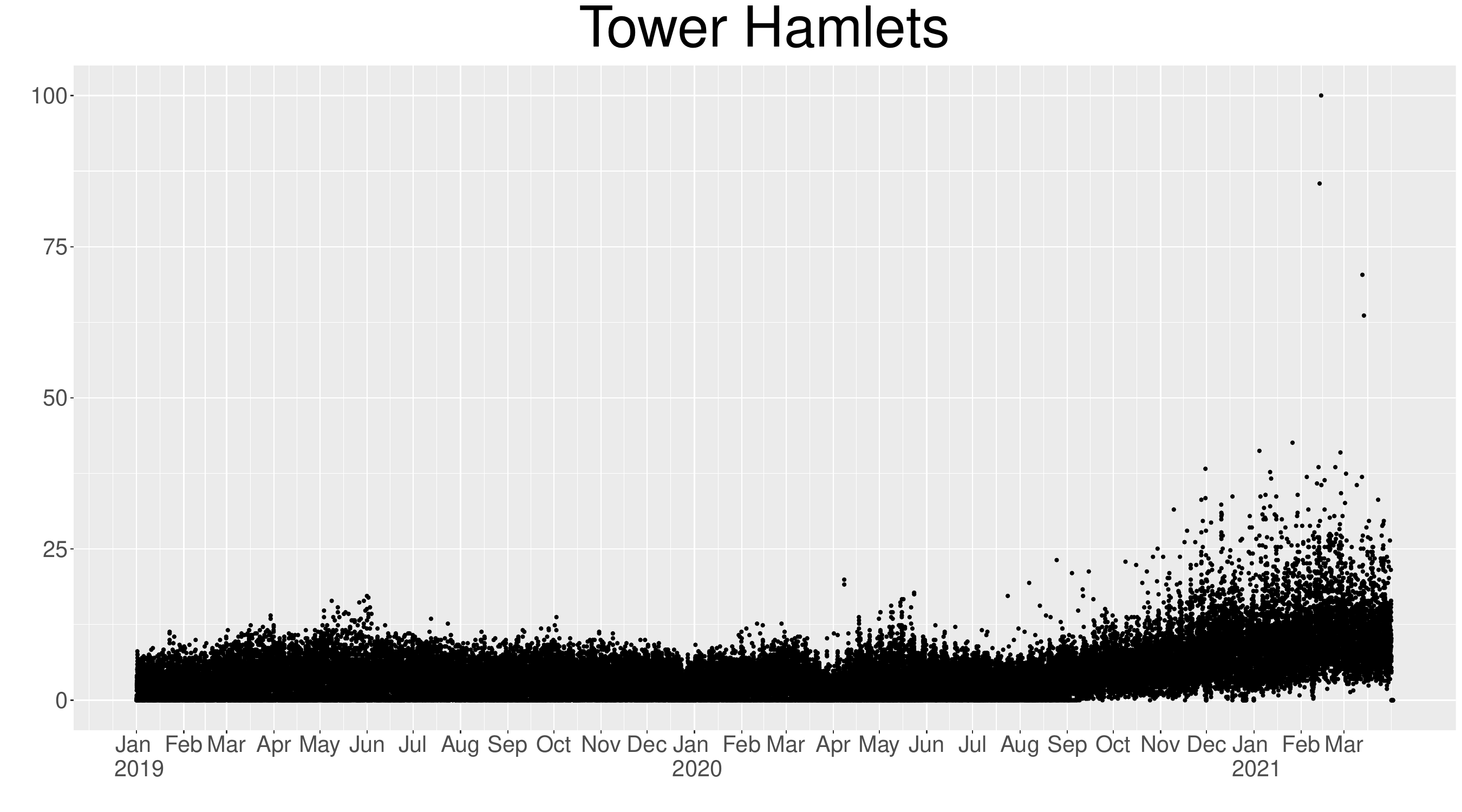}
\includegraphics[width=0.5\textwidth]{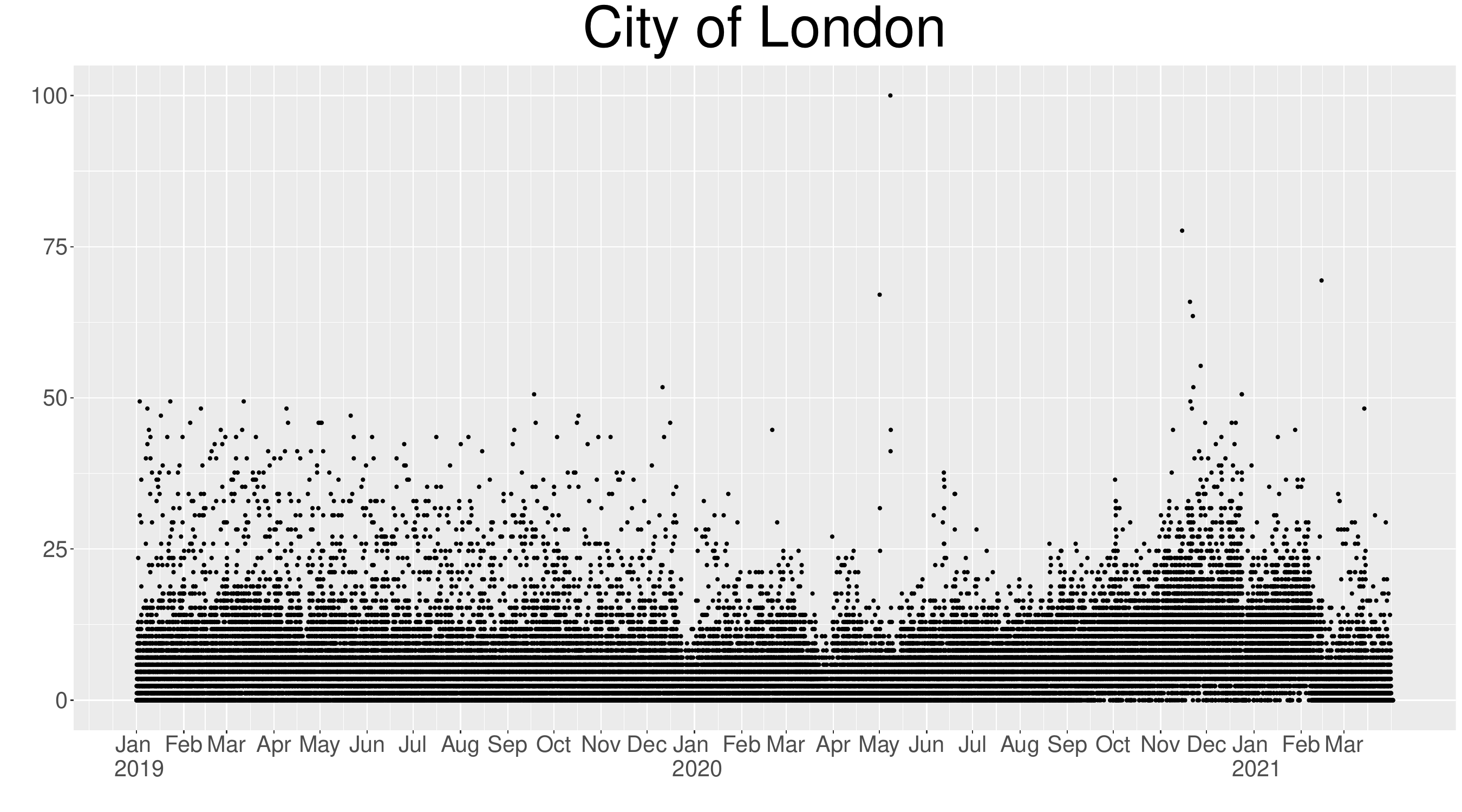}

\includegraphics[width=0.5\textwidth]{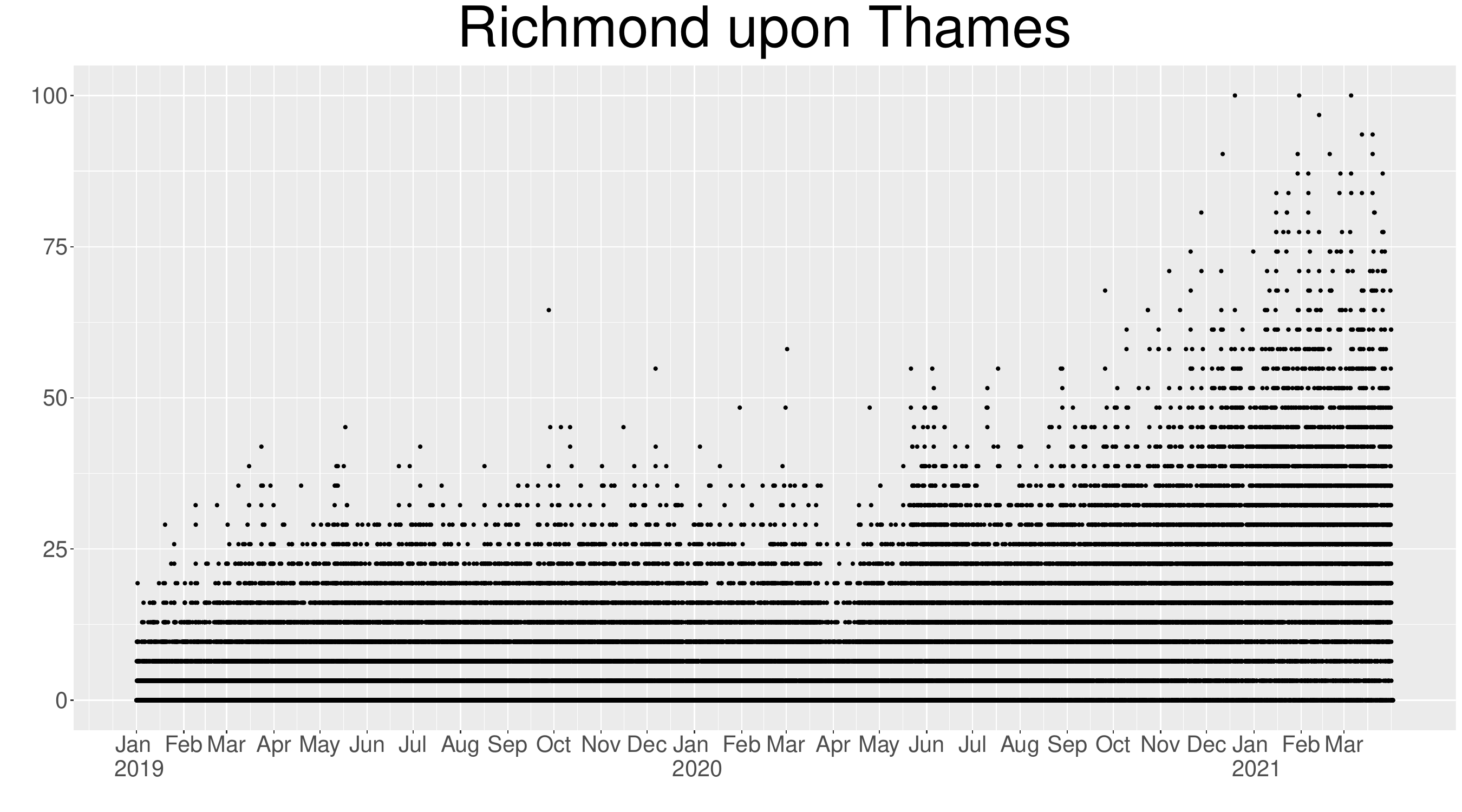}
\includegraphics[width=0.5\textwidth]{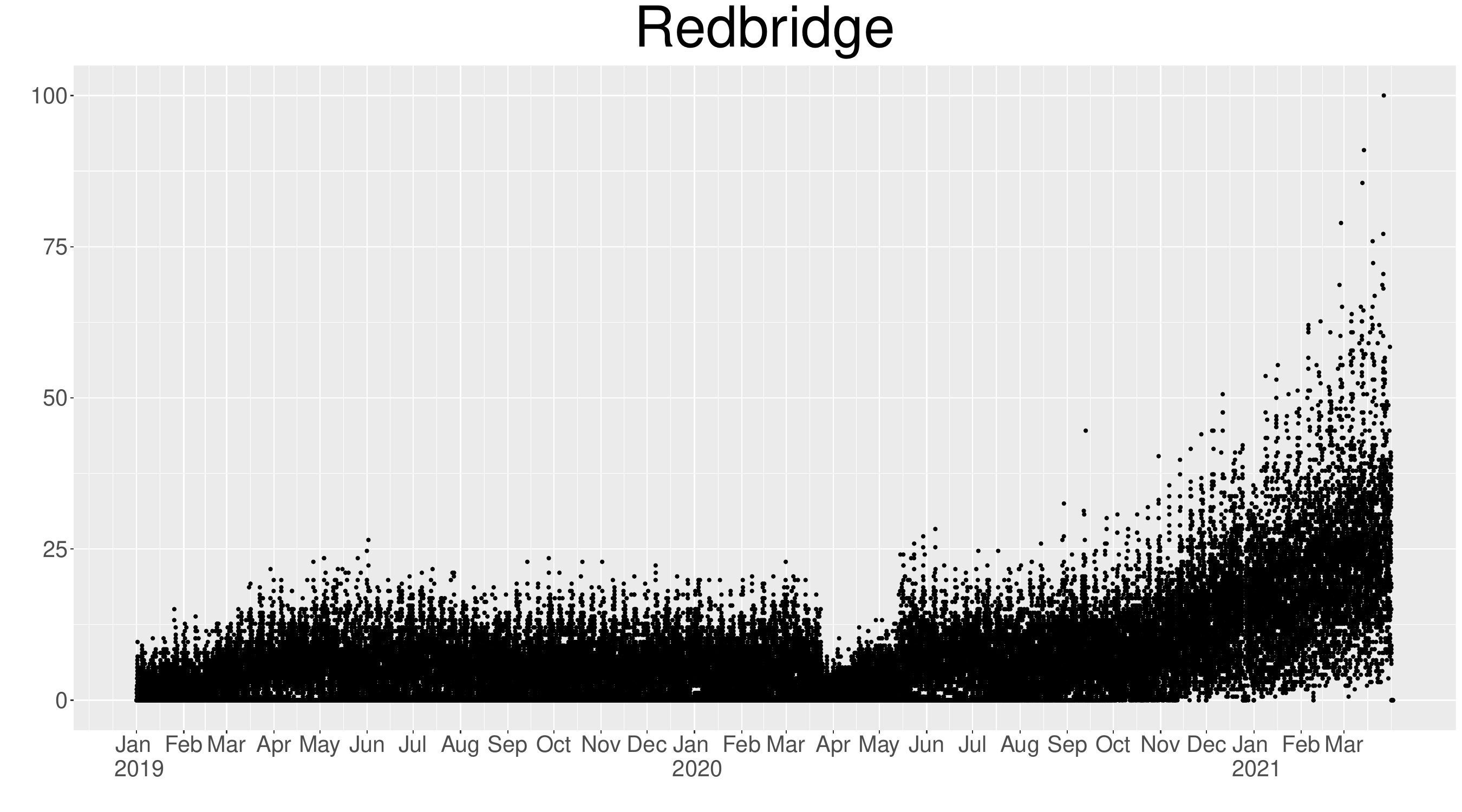}

\includegraphics[width=0.5\textwidth]{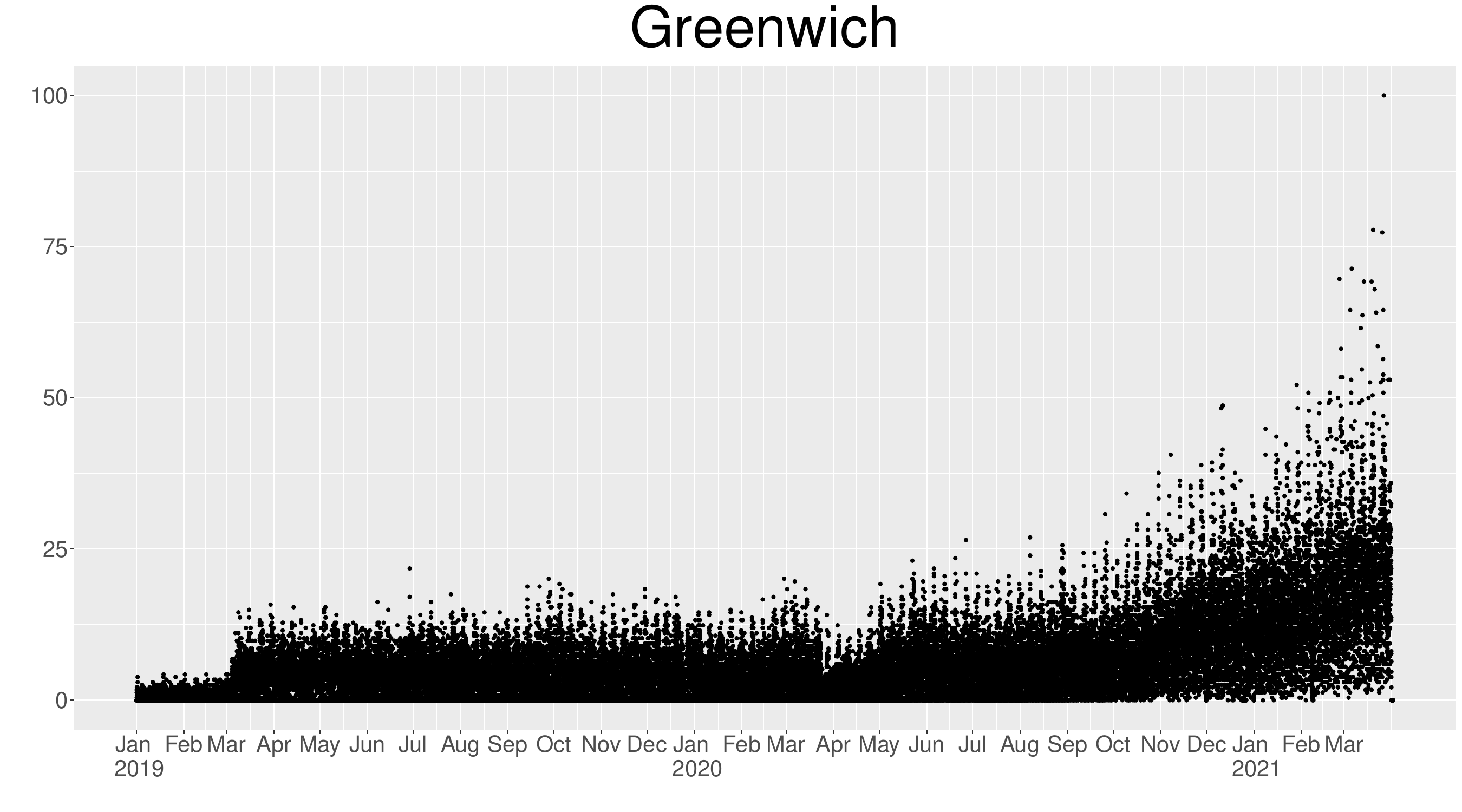}
\includegraphics[width=0.5\textwidth]{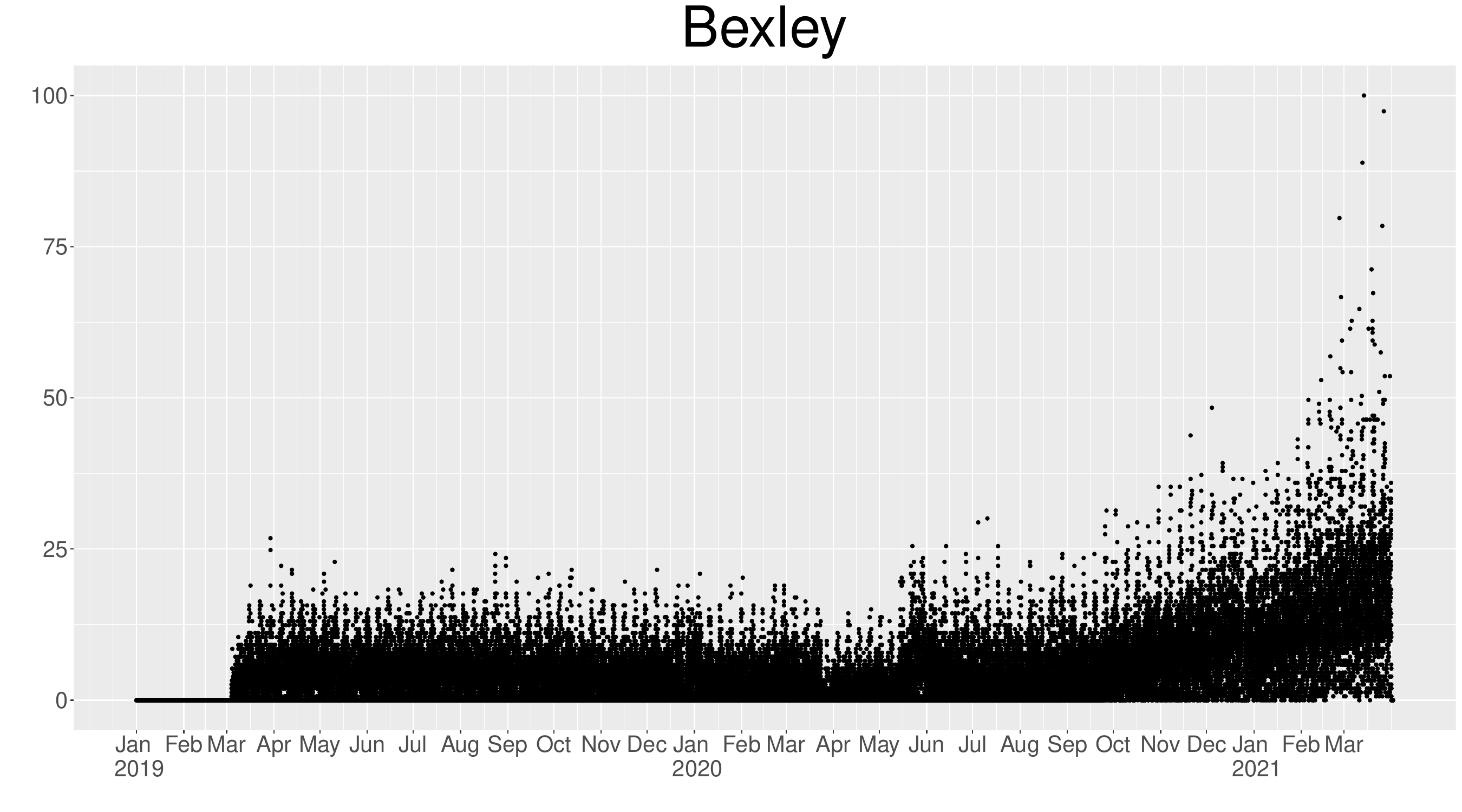}
\caption{ 15-min demand across six boroughs from January 1, 2019 until March 31, 2021. \label{fig:six_boroughs}}
\end{figure}

Figure \ref{fig:six_boroughs} plots  15-minute demand for six representative yet diverse boroughs. 
In Tower Hamlets, the borough with the highest total demand, demand is stable between (roughly) zero and ten until early 2021 when minimum demand starts increasing and becomes more volatile, thereby mounting to its highest values in the sample. 
City of London is characterized by high intra-day peak demand over the entire period requiring careful fleet planning throughout the day. 
In contrast, Richmond upon Thames has the lowest overall demand but is characterized by regular demand at specific hours.

Although visually quite different, two common distinctive features of this streaming demand data compared to more traditional time series appear. 
First,  demand is unstable. Some boroughs have low initial demand before jumping to higher levels.
The Covid-19 pandemic and first March 2020 lockdown result in a demand drop for several weeks (see Redbridge), but then demand quickly starts to grow rapidly with higher volatility until the end of the sample. 
Second, the overall range in demand is high. For example, Greenwich and Bexley start in 2019 with low variation close to zero demand and end with demand fluctuating between 0 and 100. 
This nonstationarity due to drifting levels and sudden variation of demand data necessitates an approach that can swiftly adapt by re-training ML forecast functions.  

\begin{figure}
\centering
\includegraphics[width=0.8\textwidth]{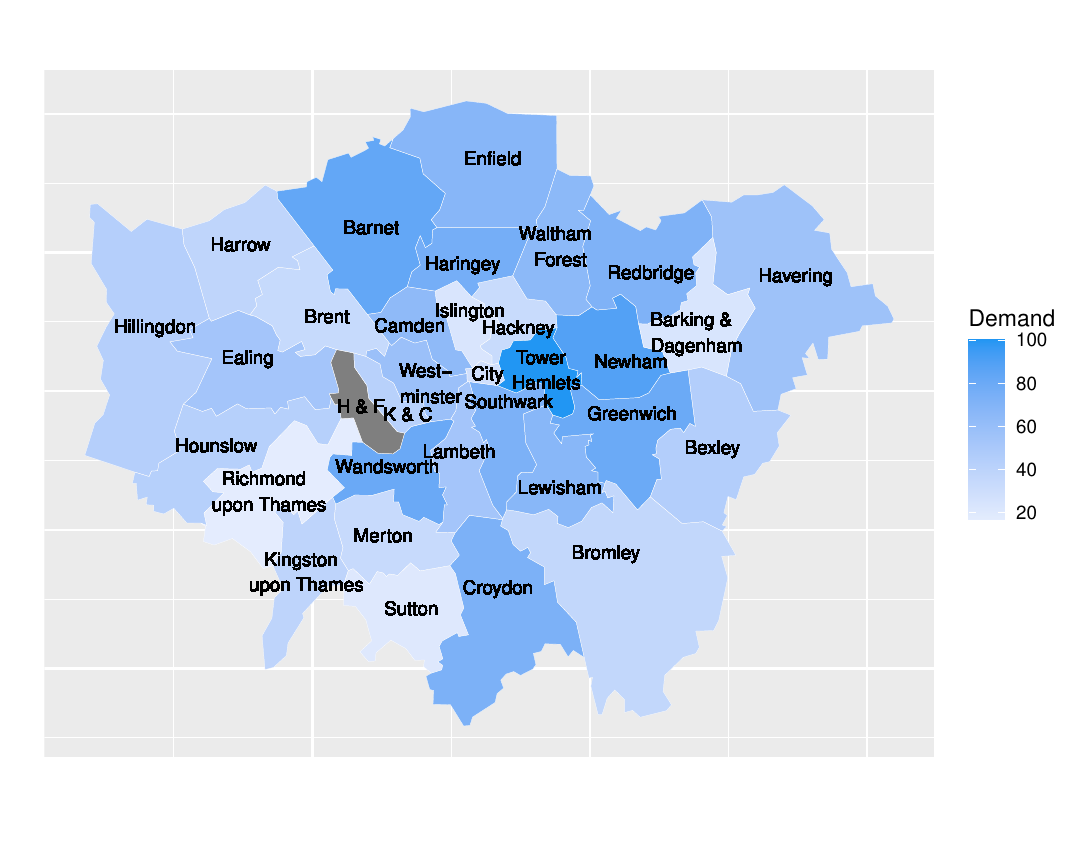}\vspace{-1cm}
\includegraphics[width=0.8\textwidth]{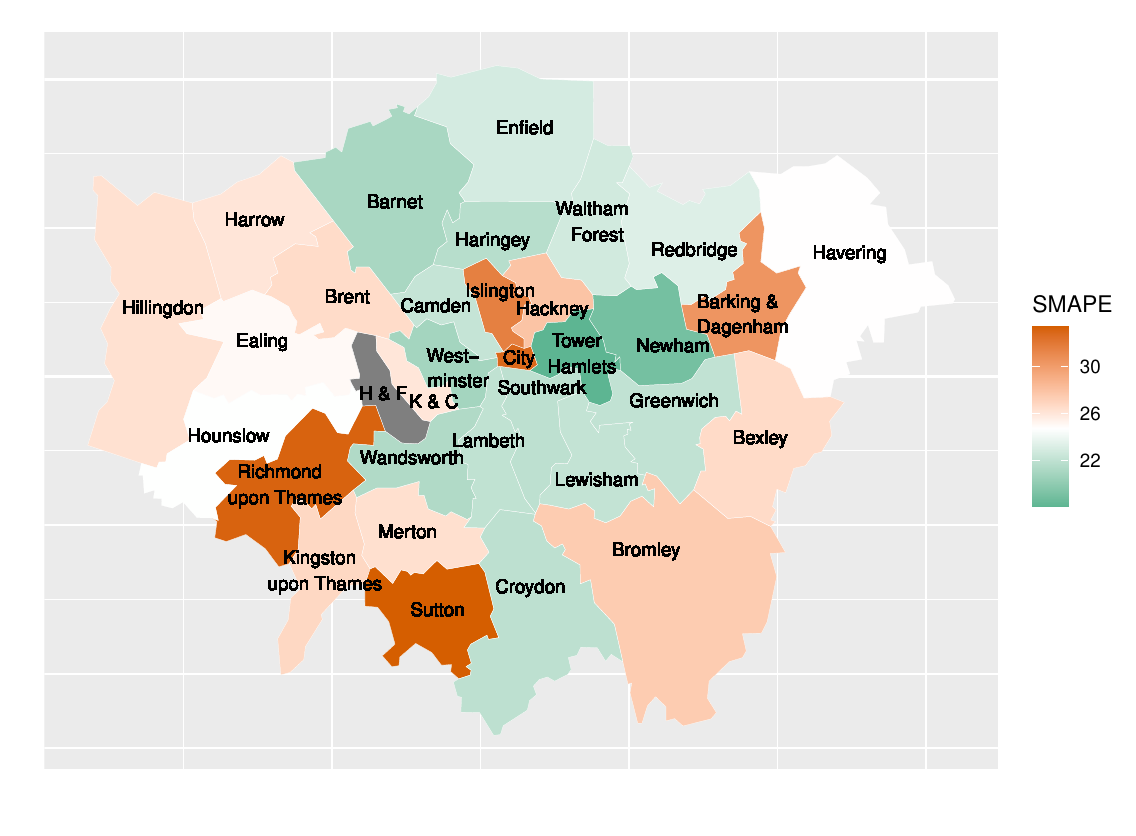}\vspace{-0.5cm}
\caption{Heat maps of demand in each borough (top: the darker the shading, the higher the demand) and forecast  errors (bottom: above, below and at median error in respectively red, green and white) for random forest monitoring, to be discussed in Section \ref{sec:forecast_results}.} \label{fig:heat_maps_london}
\end{figure}

Figure \ref{fig:heat_maps_london} (top) visualizes total demand in our sample on a London map to highlight the geographical demand differences across boroughs.\footnote{Hammersmith and Fulham (H \& F) is colored gray since no demand data is available for this borough.} Dark (light)  blue boroughs are characterized by high (low) demand. 
Demand is not uniformly distributed; the main reason for this is that fleet planning is organized with respect to drivers' availability rather than clients' demand, thereby giving rise to substantial borough-demand heterogeneity ranging from residential over business towards tourist areas.

\begin{figure}[t]
\centering
\includegraphics[width=0.49\textwidth]{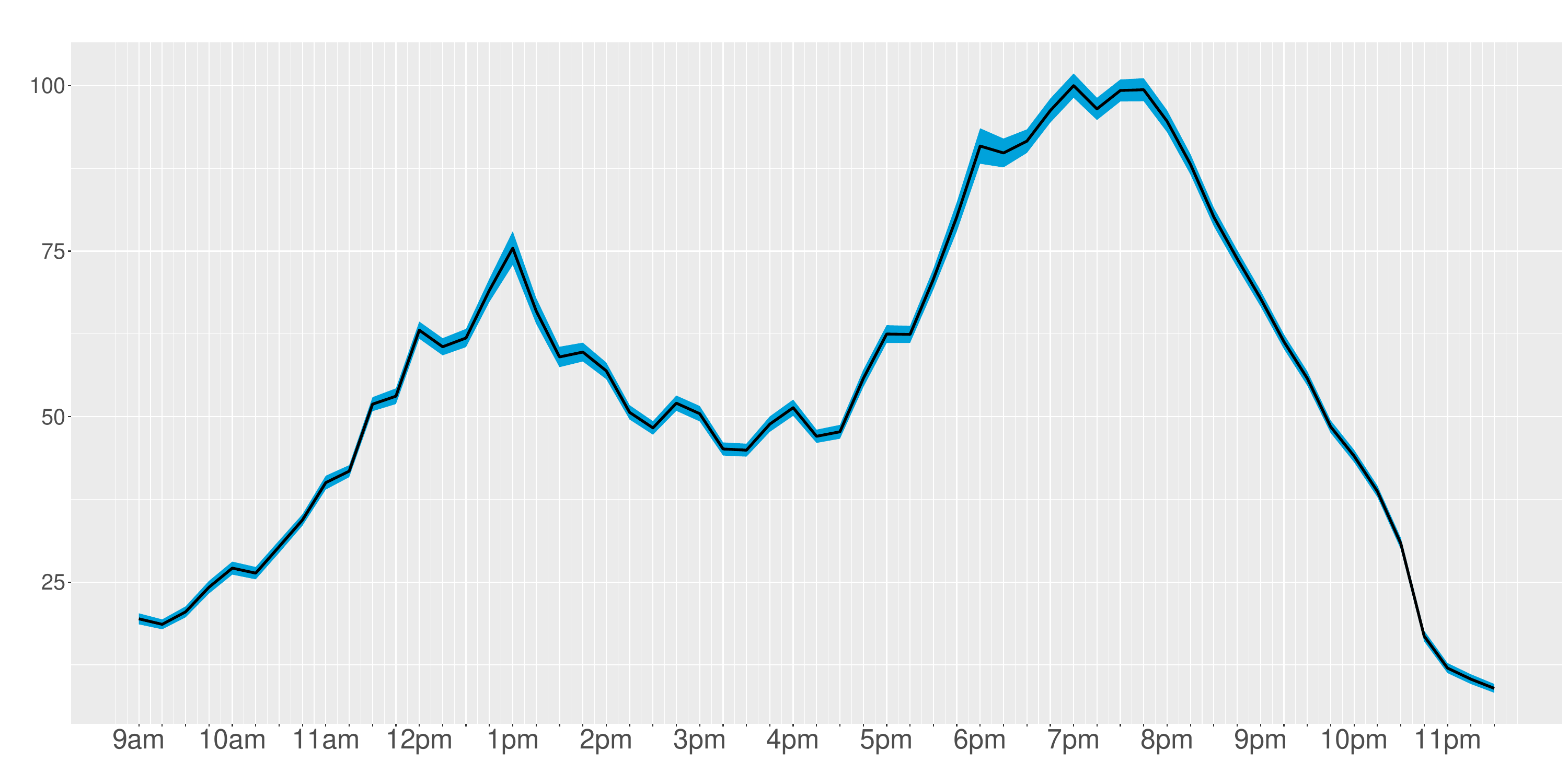}
\includegraphics[width=0.49\textwidth]{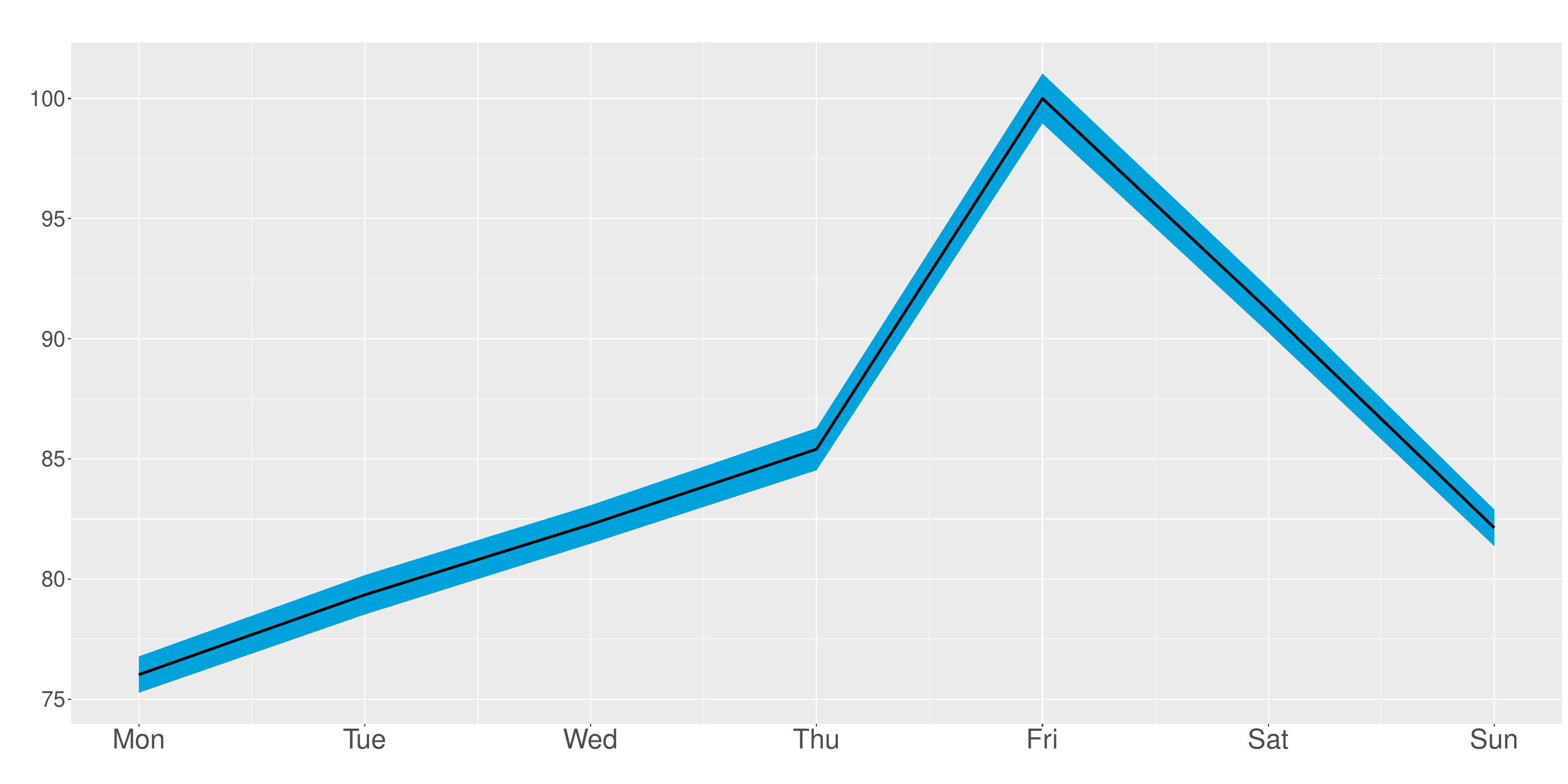}
\caption{Intra-day 15-min average hourly demand  (left) and day of the week average demand  (right) for Tower Hamlets. One standard error bands are displayed in blue. \label{fig:TowerHamlets_seasonality}}
\end{figure}

Figure \ref{fig:TowerHamlets_seasonality} (left) 
shows intra-day demand for Tower Hamlets. It becomes clear that the large majority of deliveries consists of food. In fact, demand starts low at 9am, first peaks around lunchtime between 12-2pm,  goes down only mildly between 2-5pm, but then sharply increases to peak around 7-8pm, after which it gradually decreases again to levels comparable to the morning hours. 
Besides, pronounced day of the week fluctuations are present  as well, see Figure \ref{fig:TowerHamlets_seasonality} (right). 
Average demand is lowest on Mondays, steadily increases until Thursdays, jumps on Fridays and Saturdays, and lowers on Sundays to typical Thursday levels.

\begin{figure}[t]
\centering
\includegraphics[width=0.49\textwidth]{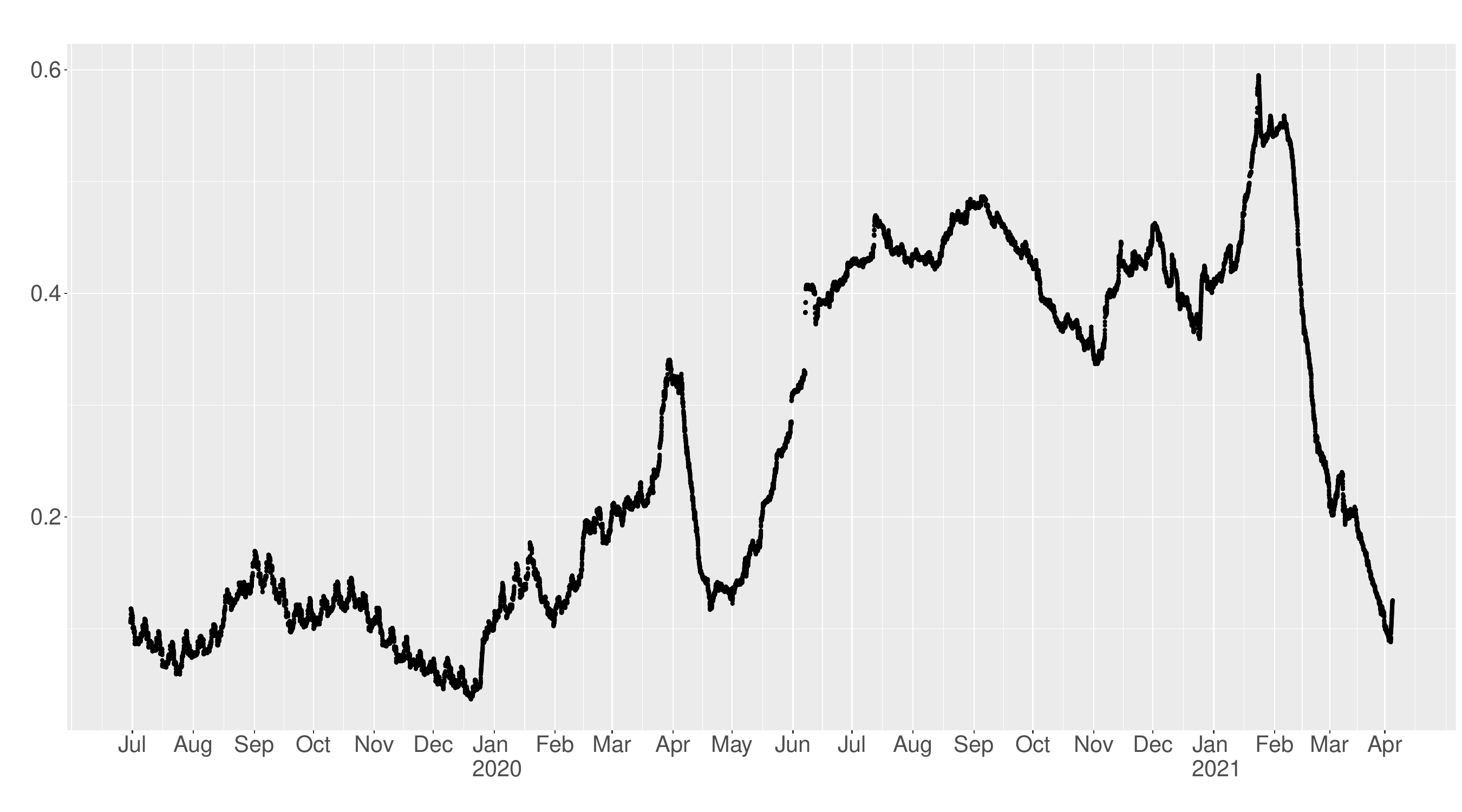}
\includegraphics[width=0.49\textwidth]{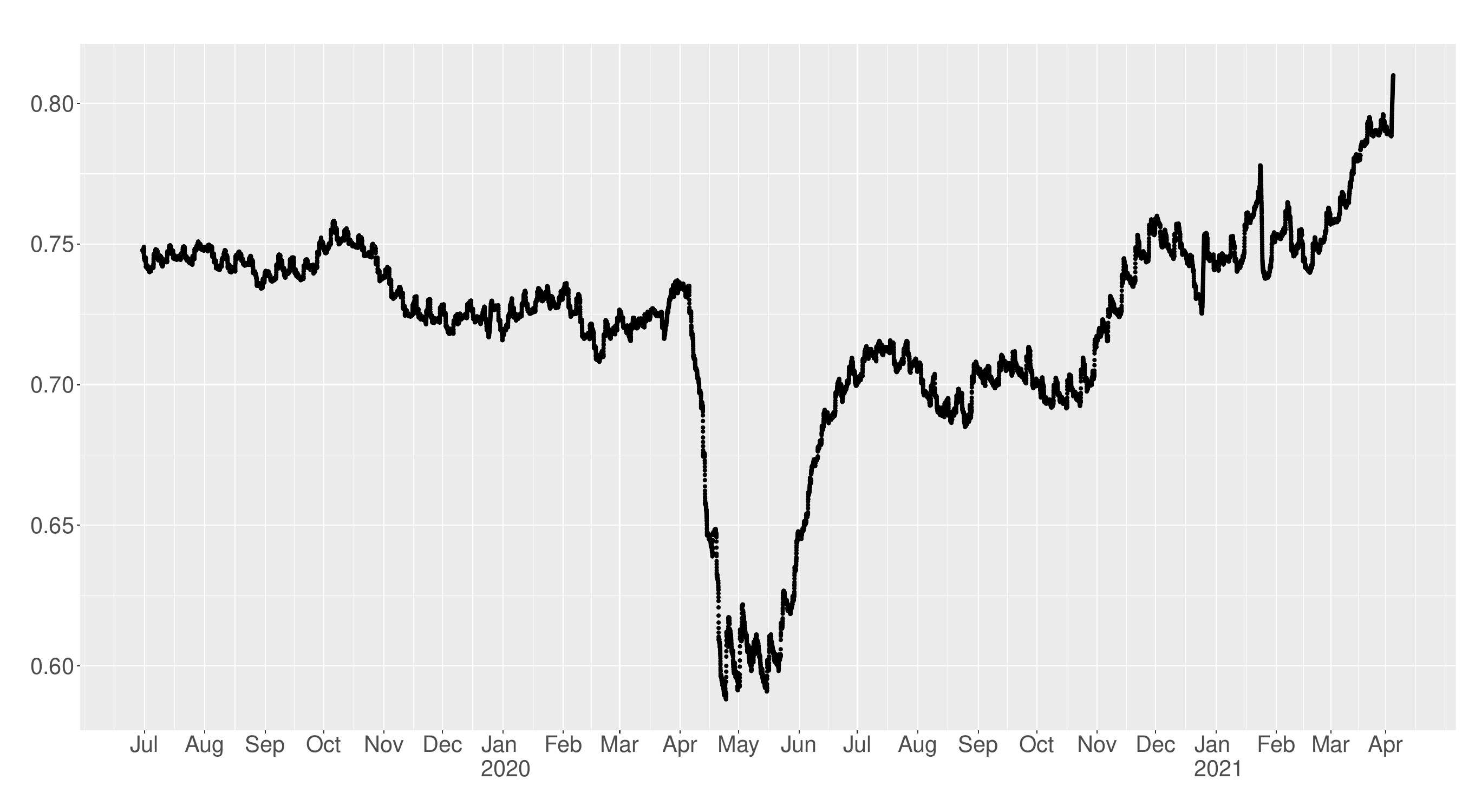}
\caption{Streaming 15-min average correlations  for City of London (left)  and Barnet (right). \label{fig:tv_correlations}}
\end{figure}

The seasonality and trend  patterns are common to several boroughs which explains their relatively high average between borough correlation of 77\% (ranging between 16\% and 91\%) over the entire sample. Correlations, however, are strongly time-varying, as can be seen from Figure \ref{fig:tv_correlations} 
which displays streaming 15-minute average correlations (for our forecast evaluation period) computed using demand from the thirty most recent days.  
City of London  has specific idiosyncratic dynamics that makes it  least related to the other boroughs with average correlations dropping below 10\%, though it has also a relatively long spell late 2020 of average correlations hovering around 40\%. In contrast, Barnet, characterized by high demand, has high average streaming correlations around 75\% throughout the sample except mid 2020 when it fell to 60\% for more than a month. The time-varying nature of these correlations makes it hard to know in advance which boroughs are predictive for which other boroughs (if any) when building a forecast model. Successful ML algorithms have to be flexible enough to capture this specific feature of our demand data.

\section{Methodology} \label{Methodology}
In Section \ref{subsec:problem} we introduce the problem of monitoring forecasting performance with streaming data.
In Section  \ref{subsec:monitoring}, we discuss our monitoring procedure.
In Sections \ref{subsec:glasso} and \ref{subsec:treeML}, we review the ML methods used to obtain the demand forecasts.

\subsection{Problem Formulation} \label{subsec:problem}
Let  $d_t = \left( d_{1t}, d_{2t}, \ldots, d_{Dt}  \right)$ 
denote the $D$-variate data stream for $t \ge 1$.  The data is non-identically distributed and dependent across time and streams. 
Data streams arrive in batches of fixed size $B$, i.e.\
the number of observations that simultaneously enter the data streams. 
A forecast model is needed to produce direct forecasts for horizons $q=1, \ldots, Q$.

We model each data stream $d_{it}$ separately conditional on all historical data streams $d_{t-j}$ for $j>1$ (as opposed to $d_{t}$ jointly conditional on $d_{t-j}$) since the monitoring procedure and corresponding re-training scheme is likely to be data-stream  (borough) specific.
A general forecast model for data stream $i$ at time $t$ that uses past demand of all streams as well as deterministic trends and seasonal effects, denoted by $a_t$, can be written  as
\begin{eqnarray}
d_{it} = f_{it}\left(\{d_{t-j}\}_{j \in \mathcal{J}}, a_t\right)  + \varepsilon_{it}, \label{eq:generalmodel}
\end{eqnarray}
where $f_{it}(\cdot)$ denotes the forecast function 
for stream $i$ at time $t$ and $\varepsilon_{it}$ represents the additive error term.

In our platform application, $t$ denotes the quarter hours and we have demand data  $d_{it}$ on $i=1,\ldots D=32$ boroughs, the batch size is $B=60$ since each business day 60 new demand values arrive for each of the quarter hours of the 15 business hours.
Forecasts are needed at the end of each business day, and this for each of the sixty quarter hours of the next business day ($Q=60)$\footnote{For ease of notation, we take the batch size and forecast horizon equal, but these can differ in applications.}. 
We include  a linear trend, hour-of-the-day and day-of-the-week seasonal dummies in $a_t$.
For the lag structure, we take
$\mathcal{J} = \{60, 420\}$, implying ``same quarter hour last day" $(d_{t-60})$ and ``same quarter hour last week" $(d_{t-420})$ lags.\footnote{From extensive experimentation with the platform demand data, we find that this lag structure allows picking up the most relevant information while keeping the dimension reasonable.} 
As a result, we use $p=86$ predictors in our platform application to forecast demand; the considered ML methods (see Section \ref{subsec:glasso} and \ref{subsec:treeML}) use the same predictor set to allow for a fair comparison.

We explicitly index the forecast function $f_{it}(\cdot)$ with $t$ to stress that our setting is nonstandard compared to the classical framework of estimating time invariant functions $f_{i}(\cdot)$. In the latter case, theory is available for both parametric and nonparametric estimators $\hat f_{i}$ that guarantee   convergence  to $f_{i}$ for increasing sample size. \cite{Delaigle_2023} propose kernel estimators for the density of data streams which smoothly vary over time. 
For streaming data and constant parametric functions, \cite{Luo_2020_renewable} show that renewable estimators have similar properties when the number of data batches goes to infinity. However, such a setup does not cover the heterogeneous data stream case  we face. In our  multivariate streaming platform data setting, we show that standard adaptations of statistical procedures do work well in practice.

A simple procedure would consist of estimating the $D$ forecast functions $f_{it}(\cdot)$ each time a new data batch of $B$ demand values becomes available.
Such routine is, however, computationally too expensive for an on-demand platform to implement in production, as they collect  data at high-frequency intervals for a market consisting of many delivery areas (large $D$). 
To keep the production of streaming forecasts feasible, we therefore suggest  a monitoring scheme that determines when re-training is needed, as detailed next.

\subsection{Monitoring Forecast Loss Streams} \label{subsec:monitoring}
Instead of re-estimating the function $f_{it}(\cdot)$ for each data stream at the baseline frequency $t$, 
we wish to keep the forecast functions mainly constant, as regulated via two  dynamics: one related to the batch structure of the data and one related to the detection of forecast instability.

First, taking the batch structure of the data into account, we only consider re-training of the forecasting function whenever a new data batch arrives, so at (end of batch) times denoted by subindices $b : =\{t: \text{mod}(t,B)=0\}$, where mod stands for modulus. Hence, we assume that  the forecast function is constant intra-batch: 
\begin{eqnarray}
{f}_{ib} =  f_{ib+1} = \ldots = f_{ib+B-1} \quad \forall b.
\end{eqnarray}

Second, at each end of batch time $b$, we determine whether the forecast function should be re-estimated or not. 
We  keep the forecast function constant (hence no re-training) under ``forecast stability". To define the latter, we introduce additional notation. 
Denote the forecasts made at batch-end $b$ for data stream $i$ at horizon $q=1,\ldots, Q$ by 
\begin{eqnarray}
 d_{ib+q|b}^f = \hat{f}_{ib}(\{d_{b+q-j}\}_{j \in \mathcal{J}}, a_{b+q}),   
\end{eqnarray}
and let 
\begin{eqnarray}
    l_{ib+q}=l(d_{ib+q|b}^f,d_{ib+q})
\end{eqnarray}
be the forecast loss, with $l(\cdot,\cdot)$ any loss function computing forecast errors. 
At the end of the next batch period $b+B$, we compute these forecast losses and investigate whether the forecast loss  for data stream $i$ given by
\begin{eqnarray}
   L_{ib+B}=\{l_{ib+q}\}_{q=1}^Q 
\end{eqnarray}
is significantly different, on average, from past losses in a reference batch denoted by $L_{ib}^0$.

Under the described monitoring  procedure, the null hypothesis at time $b+B$ is 
\begin{eqnarray}
    \mbox{H}_0:  \mbox{E} (L_{ib}^0) = \mbox{E} (L_{ib+B}),
\end{eqnarray}
which we test against a two-sided alternative with a standard equality of means test.  In case we reject the null hypothesis for stream $i$,  there is ``forecast instability", we re-estimate the forecasting function, and re-initialize the reference batch.
In case we do not reject the null hypothesis, 
there is ``forecast stability", we keep the forecast function constant and the losses become part of the reference batch.
Note that the execution of the standard test requires computing only sample moments and is therefore extremely fast. Therefore, monitoring can be implemented in a production setting for an indefinite amount of iterations. In the application, we test forecast stability every end of business day using forecast losses computed over 60 quarter hours and the reference batch losses.

In sum, our monitoring procedure  re-estimates the  forecast functions $f_{it}$ for each stream $i$ at different, a priori unknown times. The time points at which re-training is triggered can be summarized via the streaming ``re-training" indicator  $r_{it}$ defined as 
\begin{equation*}
r_{it} = \begin{cases}
0 & \quad \text{mod}(t,B) \neq 0 \; \text{OR} \; \left(\text{mod}(t,B)=0 \; \text{AND} \;\text{forecast stability}\right), \\
1 & \quad \text{mod}(t,B)=0 \; \text{AND} \;\text{forecast instability}.
\end{cases}
\end{equation*}
A  zero value simply indicates a time point of forecast stability of stream $i$, whereas a value of one indicates a time point of forecast instability and hence the necessity to re-estimate the forecast function. The stream $r_t = \left( r_{1t}, r_{2t}, \ldots, r_{Dt}  \right)$ can thus be used for stability tracking. Algorithm \ref{alg:monitor} gives an overview of the  flow of the proposed monitoring procedure.

\begin{algorithm}[t]
\caption{\color{black} Monitoring ML Forecast Streams}
\label{alg:monitor}
\leading{15pt} 
\begin{algorithmic}[1]
\STATE Let $S_{ib}$   be initial data batches for $b=0$ and $i=1,\ldots, D$ where $D$ is the number of data streams
\STATE Estimate forecast functions $ f_{i0} $ on $S_{i0}$, ($i=1,\ldots, D$),  compute initial  forecasts 
\STATE  Read data batch $S_{i1}$, compute forecast losses to obtain  initial reference batch $L_{i1}^0$
\FOR{$b = 2, 3, \ldots$}

	\FOR{$i = 1$ to $D$}
	  	\STATE Read data batch $S_{ib}$ 
	  	\STATE Compute forecast losses $L_{ib+B}=\{l_{ib+q}\}_{q=1}^Q$ 
	  	\STATE Test for equal average forecast loss with reference batch  $L_{ib}^0$ resulting in $TEST=1 (0)$ if reject (accept)
		\IF{$TEST=1$} 
              \STATE Estimate forecast functions $ f_{ib}$
			 \STATE Reset reference batch $L_{ib}^0$ and set $r_{ib}=1$
            \ELSE
                 \STATE Update reference batch $L_{ib}^0$ with $L_{ib+B}$ and set $r_{ib}=0$
		\ENDIF    
    	  \STATE Compute forecasts $\hat{f}_{ib}(\{d_{b+q-j}\}_{j \in \mathcal{J}}, a_{b+q})$
	\ENDFOR
\RETURN $D$ forecasts for decision making	
\ENDFOR
\end{algorithmic}
\end{algorithm}

The size properties of a standard sequentially implemented test in the case of our non-identically distributed data stream implied loss functions are difficult to assess.\footnote{Recently, research on two sample tests has been mostly focusing on  the high dimensional iid data case, e.g., \cite{Zhang_JASA_2020} and \cite{Jiang_JASA_2022}.} 
However, from our experience, given that the reference batch $L_{ib}^0$ is frequently reset for platform data streams, and thus also the null hypothesis, and given that the sample size of $L_{ib}$ is  $Q=60$, we expect reasonable size properties of our test. This is confirmed in Table \ref{tab:MonteCarlo} which presents simulation results under the null hypothesis of forecast stability. For Gaussian data streams with lengths ranging between 10,000 and 100,000, and data batch sizes from 10 to 100 observations, the empirical rejection frequencies are only slightly higher than their nominal levels. For Chi-square (with five degrees of freedom) data streams the results are  similar.

\begin{table}[t]
\begin{center}
\caption{Rejection frequencies under forecast stability. \label{tab:MonteCarlo}}
\begin{tabular}{lccccccc}
  \hline
\multicolumn{8}{c}{Gaussian Data} \\[0.2cm]  
 Length stream  &  \multicolumn{3}{c}{5\% significance level} && \multicolumn{3}{c}{1\% significance level}  \\[0.1cm] \hline 
  &  \multicolumn{3}{c}{Batch size} && \multicolumn{3}{c}{Batch size}  \\[0.1cm] 
   & 10 & 50 & 100 &&  10 & 50 & 100  \\ 
     \cline{2-4}  \cline{6-8}\\[-0.4cm]

10,000 &0.067	&	0.070	&	0.071	&&	0.013	&	0.013	&	0.013	\\
20,000 &0.067	&	0.071	&	0.072	&&	0.013	&	0.013	&	0.013	\\
50,000 & 0.067	&	0.071	&	0.072	&&	0.013	&	0.013	&	0.014	\\
100,000 &0.067	&	0.071	&	0.072	&&	0.013	&	0.013	&	0.014	\\
   \hline
\multicolumn{8}{c}{ } \\[-0.1cm]  
\multicolumn{8}{c}{Chisquare(5) Data} \\[0.2cm]  
 Length stream  &  \multicolumn{3}{c}{5\% significance level} && \multicolumn{3}{c}{1\% significance level}  \\[0.1cm] \hline 
  &  \multicolumn{3}{c}{Batch size} && \multicolumn{3}{c}{Batch size}  \\[0.1cm] 
   & 10 & 50 & 100 &&  10 & 50 & 100  \\ 
     \cline{2-4}  \cline{6-8}\\[-0.4cm]

10,000 &0.074	&	0.073	&	0.072	&&	0.022	&	0.016	&	0.015 \\
20,000 &0.074	&	0.072	&	0.071	&&	0.022	&	0.016	&	0.015 \\
50,000 &0.073	&	0.073	&	0.072	&&	0.022	&	0.016	&	0.015 \\
100,000 &0.073	&	0.073	&	0.072	&&	0.022	&	0.017	&	0.015 \\
\hline
\end{tabular}
\end{center}
\raggedright	
\footnotesize
Notes: Rejection frequencies of the monitoring test under the null hypothesis of forecast stability. Batch size indicates the number of observations that are sequentially  added to the  stream.  The top (bottom) panel reports results for streams from Gaussian (Chi-square, with 5 degrees of freedom) data.
The results are based on 1,000 replications by running Algorithm \ref{alg:monitor}.  
\end{table}

In the case of homogeneous data streams modeled with linear regressions, \cite{White_ECA_1996} illustrate that standard tests reject too often.
For our monitoring purposes, an oversized test implies more frequent re-training, at the benefit of forecast accuracy but the cost of computing time.  Our approach resembles \cite{Luo_JASA_2022_batches} who monitor abnormal data batches in data streams modeled with time invariant parametric functions. Using the  standard test of \cite{Hansen_Econometrica_1982}, their null hypothesis regards the equality of model parameters using score vectors between a fixed reference batch and a new incoming batch. In our setup, the forecast loss function streams are only temporarily in a stable regime, which explains why we define a new reference batch at each date with detected forecast instability. 

To apply our monitoring procedure in practice, we need to estimate the forecast function $f_{it}(\cdot)$. While our monitoring procedure works equally well with any forecast method, we consider regression-based (Section \ref{subsec:glasso}) and tree-based (Section \ref{subsec:treeML}) ML methods.

\subsection{Linear Regression-based Machine Learning} \label{subsec:glasso}
The easiest way to characterize $f_{it}(\cdot)$ explicitly is through a linear regression model.
Given the expert knowledge from the logistics platform, such a ``surface plus error" model can be competitive against the ML procedures, see \cite{Efron_survey_2020} for a recent discussion.

The linear regression model can be written in matrix form as $y =  X \beta + \varepsilon$ where 
$y$ contains the to be forecast demand,
$X$ collects all $p=86$ predictors namely the deterministic components (trend and seasonal dummies) as well as the lagged demand values (see equation \eqref{eq:generalmodel}), and 
$\beta$ is the parameter vector.  
As popular parametric, regression-based  ML method, 
we consider the lasso \citep{tibshirani1996regression}.
The lasso estimator is obtained by minimizing 
$$
\frac{1}{T} \sum_{t=1}^{T} (y_t - x_t^\top \beta)^2 + \lambda ||\beta||_1,
$$
where $\lambda>0$  is a tuning parameter ($\lambda=0$ corresponds to Ordinary Least Squares) and $||\beta||_1$ denotes the $\ell_1$-norm of the parameter vector $\beta$.
By adding the $\ell_1$-norm to the objective function, a sparse parameter estimate is obtained where several elements of $\hat{\beta}$ are set to zero.
The larger the value of the tuning parameter, the sparser the obtained estimate.
We refer the interested reader to \citeauthor{hastie2009elements} (Section 3.4.2, \citeyear{hastie2009elements}) for textbook explanations.
We use the \texttt{glmnet} package \citep{glmnet2010, glmnet} in \verb|R| \citep{Rcoreteam}   to obtain the lasso; the tuning parameter is selected via the Bayesian Information Criterion.

The lasso takes linearly into account lagged demand of all considered boroughs and seasonal effects to forecast borough demand. The  model  is fully parametric and the estimates will only reflect linear correlation patterns in the data.
 However,  how exactly boroughs interact might be more complex than the linear relationships specified above. Thus, we introduce additional flexibility by  implementing two tree-based ML algorithms-- described in the next section --that have proven their success in many applications. Similar to the baseline linear model, both algorithms use $y$ as target 
and $X$ as predictor/feature matrix for training.

\subsection{Tree-based Machine Learning}\label{subsec:treeML}
While our procedure allows one to use their (non-parametric) ML method of preference, we deliberately opt for tree-based methods over neural networks or deep learning methods.
The reason for this choice  aligns well with the discussion recently provided in \cite{januschowski2022forecasting}.
For tree-based methods, such as random forecast and XGBoost, robust and sophisticated  implementations with sensible default parametrization are now routinely available across standard software packages such as \texttt{R} or Python, making their ``default'' solutions often competitive across a variety of tasks.
Besides, they regularly appear as top contestants of forecast competitions  (M4 and M5 competitions, the global energy forecasting competitions, and various Kaggle forecasting competitions discussed in \citealp{bojer2021kaggle}) where, importantly, ``off-the-shelf' implementations are used without modifications to the software or core model. 
In contrast, neural networks and deep learning methods have not yet reached the same robustness in terms of their implementation, are often time-consuming to train and typically require custom-made code to achieve competitive performance.
When time is short as for digital platforms, the ability to rely on a mature model implementation with sensible defaults is key, thereby motivating our choice for tree-based ML methods, namely random forest and XGBoost.
\medskip

\noindent
{\bf Random Forest.} 
A random forest estimates $f_{it}(\cdot)$ by combining regression trees. A regression tree is a nonparametric method  that partitions the feature space, i.e.\ in our case lagged demand, trend and seasonality effects, to compute local averages as forecasts, see \cite{Efron_Hastie_2016} for textbook explanations. The random forest tuning parameters are the number of trees that are used in the forecast combination, the number of features to randomly select when constructing each regression tree split, and the minimum number of observations in each terminal node to compute the local forecasts. 
Since default settings often attain excellent performance across a variety of settings, 
 we use the standard implementation of the \texttt{randomForest} package \citep{randomForest} in \verb|R|.
Methodologically, random forests are proposed by \cite{Breiman2001}, consistency and asymptotic normality results are put forward by respectively \cite{Scornet_2015} and \cite{Wager_Athey_2018} for iid data.  Empirically, random forests have shown good forecasting performance  in several applications, see  for example \cite{Medeiros_2021} for  a recent application on inflation forecasting.
\medskip

\noindent
{\bf XGBoost.} 
Gradient boosting, proposed by \cite{Friedman_AOS_2001}, constructs the forecasting function $f_{it}(\cdot)$  by sequentially fitting small regression trees, i.e.\ weak learners, to the residuals by the ensemble of the previous trees. This procedure results in a one final tree, constructed as a sum of trees, used for forecasting in contrast to the random forest where the forecast results as the average of many trees. The properties of gradient boosting have been well studied for iid data, see for example \cite{Buhlmann_2006} and \cite{Bartlett_2007} and the references therein.
Extreme gradient boosting (XGBoost) is introduced by \cite{Chen_XGBoost_2016} and is an algorithm that optimizes the implementation of the gradient boosting framework in terms of speed and flexibility. The popularity of XGBoost among data scientists is large. There are several tuning parameters that can strongly impact the performance, and cross-validation techniques can be used to optimize them. However, to avoid prohibitive computational cost in our streaming data setting, we fix them as follows: 100 boosting iterations, 6 as max tree depth,  0.3 as amount of shrinkage,
0 as minimum loss reduction, 1 as subsample ratio of columns, 1 as minimum sum of instance weight, 1 as subsample percentage. These values are in line with \cite{He_JASA_2021} and practitioners.  We use the 
\texttt{xgboost} package \citep{chen2015xgboost} in \verb|R|.  

\section{Forecast Setup,  Benchmarks and Forecast Metric} \label{Forecasting}
We detail our  set-up to assess the forecast performance of our monitoring procedure, describe the benchmarks against which its performance is compared, and we motivate the forecast error metric we use to facilitate comparison of forecast performance across boroughs.

We implement the following monitoring forecasting procedure for each of the 32 boroughs in London. 
To ensure data availability across all boroughs, we use data as of March 1, 2019 and start training  each ML algorithm from an initial sample period of 180 days of 15-minute data. At business day-end, we produce forecasts for quarter hour demand  $d_{ib+q|b}^f$ (from 9am to 11pm) for the next day, i.e.\ a forecast horizon $q=1,\ldots, B=60$.\footnote{Since we forecast all quarter hours for
the next day, the forecast horizon $Q$ and batch size $B$ are equal (namely 60) in our application. To facilitate readability, we simply use $B$ to denote both in this section.} The out-of-sample forecast evaluation period thus runs from August 28, 2019 to March 31, 2021. 
This generates a daily stream of quarter hour forecasts that are evaluated against  actual demand  to obtain the average daily loss batch $\frac{1}{B} \sum_{q=1}^B (d_{ib+q} - d_{ib+q|b}^f)^2$.  If this batch is significantly different from the reference batch, then the proposed monitoring procedure triggers re-training of the ML algorithm. 
To implement the standard equality of means test, we use  the function \texttt{t.test} (with default arguments) in \verb|R| and set the size of the test at five percent.
The last 180 days of data are then used to update the new forecast function $\hat f_{it}$. If the monitoring test does not reject, then forecasts are computed with the existing forecast function. We 
re-train an ML algorithm on the most recent 180 days (hence a rolling window) in order to keep the computation time constant.\footnote{We compared this with an expanding window approach, i.e.\ keeping all historical data when re-training. The main insights were very similar (i.e.\ the relative forecast performance of the methods remained unaltered) but the computational burden was considerably higher.}

The proposed monitoring procedure has the appealing feature of checking for forecast instability via standard two sample mean tests, which are implemented in any data software and easily understandable for  analysts. We also assess the performance of the monitoring procedure where instead 
a standard, off-the-shelf-available changepoint test on the daily streaming forecast loss is used as viable alternative. To this end, we use the computationally efficient, easy-to-use mainstream Pruned Exact Linear Time (Pelt) algorithm of \cite{killick2012optimal} in the \verb|R|  package \verb|changepoint| \citep{killick2014changepoint}, with a normal likelihood test that allows for changes in both mean and variance. 
Many alternative online changepoint tests exist, see for example FOCuS of \cite{Fearnhead_JMLR_2023} for univariate streams or OCD method of \cite{Chen_Wang_JRSSB_2022}
for multivariate streams.
Since we monitor univariate data streams, we also compared the performance of the offline Pelt procedure to the online FOCuS procedure (see Table \ref{Table-Pelt-vs-Focus} in  Appendix \ref{app:results}), but only report results on Pelt in Section \ref{sec:forecast_results} since it is, overall, the most competitive benchmark.

Additionally, we compare the performance of the 
monitoring procedures with the following benchmarks. The first benchmark is simply $d_{ib+q|b}^f =  d_{ib+q-420}$, which we call ``Naive"  since  it simply implies that the forecast of a specific quarter hour is the value observed in the same quarter hour and day of the  previous week. This way of forecasting is popular in the industry since it is ultra fast, i.e.\ it does not require parameter estimation, and it works well in the case of strong seasonality patterns as observed in platform  data. 
The other two benchmarks re-train the ML algorithms for all boroughs at deterministic time intervals in contrast to our monitoring approach that triggers re-training at prior unknown times that may vary across boroughs. 
We either update the forecast functions very fast, namely daily, or only very 
slowly, namely every semester, and  in both cases we do this for each borough-- which corresponds to the unrealistic situation of a joint borough forecast instability. 
The deterministic daily benchmark is computationally unfeasible in the live streaming data environment for which we develop our monitoring approach. Nevertheless, despite the one-time heavy computational cost to report results in this paper, such a  benchmark allows contrasting how much forecast performance is lost by re-training only in case of instability.

To evaluate  out-of-sample forecast accuracy, we use the symmetric absolute percentage error loss  $L^{\mbox{sape}} (d_{ib+q},d_{ib+q|b}^f)= 100 \mid d_{ib+q} - d_{ib+q|b}^f \mid / (|d_{ib+q}| + |d_{ib+q|b}^f|)$,  a popular  metric at delivery platforms because of its
(i)  in-built asymmetry where under-forecasting (forecast below actual) is penalized more than over-forecasting (forecast above actual) and (ii) relative nature which facilitates comparisons across heterogeneous boroughs. We average $L^{\mbox{sape}} (d_{ib+q},d_{ib+q|b}^f)$ across the out-of-sample quarter hours and denote this as the SMAPE.

\section{Results} \label{sec:forecast_results}
This section is divided in three parts. 
In Section \ref{subsec:results:forecast}, we discuss the forecast performance of the updating procedures-- monitoring, naive and deterministic benchmarks --and the three ML methods-- random forest, XGBoost, lasso --across all boroughs as measured by the SMAPE.
In Section \ref{subsec:results:monitor}, we gather insights regarding our monitoring procedure in terms of frequency and timing of the detected instabilities. 
In Section \ref{subsec:results:shap}, we interpret which parts of the information set generate predictability via the case study of City of London.

\subsection{Overall Forecast Performance Across Boroughs} \label{subsec:results:forecast}

\begin{table}[t]
\caption{SMAPEs for all boroughs across the the updating procedures and ML methods. \label{tab:SMAPE_all_boroughs}}
\resizebox{0.68\textwidth}{!}{\begin{minipage}{\textwidth}
\begin{tabular}{lccccccccccccc}
  \hline
Borough &  \multicolumn{4}{c}{Random Forest} & \multicolumn{4}{c}{XGBoost} & \multicolumn{4}{c}{Lasso} & Naive  \\[0.1cm] 
   & \multicolumn{2}{c}{Monitoring} & \multicolumn{2}{c}{Deterministic} & \multicolumn{2}{c}{Monitoring} & \multicolumn{2}{c}{Deterministic} &  \multicolumn{2}{c}{Monitoring} &\multicolumn{2}{c}{Deterministic} &  \\ 
   & Ours & Pelt & daily & semester & Ours & Pelt  & daily & semester & Ours & Pelt  & daily & semester & \\[0.1cm]  \hline
TowerHamlets & 18.07 & 21.39 & 16.20 & 23.94 & 20.02 & 23.39 & 17.91 & 28.28 & 19.47 & 22.40 & 17.62 & 26.26 & 20.34 \\ 
  Wandsworth & 21.38 & 23.44 & 19.17 & 27.55 & 23.51 & 28.21 & 20.60 & 29.04 & 22.60 & 26.81 & 20.40 & 29.20 & 23.58 \\ 
  Camden & 22.25 & 25.61 & 20.66 & 28.76 & 24.90 & 30.24 & 22.10 & 29.07 & 23.46 & 25.07 & 21.65 & 29.47 & 25.77 \\ 
  Islington & 31.55 & 33.85 & 30.04 & 36.67 & 34.27 & 35.39 & 31.96 & 39.31 & 32.88 & 34.59 & 31.55 & 38.11 & 37.69 \\ 
  Westminster & 20.87 & 22.52 & 19.22 & 25.32 & 22.72 & 28.02 & 20.77 & 27.16 & 22.53 & 24.35 & 21.12 & 27.15 & 24.21 \\ 
  Lambeth & 22.01 & 24.75 & 20.02 & 24.54 & 24.02 & 29.89 & 21.82 & 26.50 & 23.63 & 25.31 & 21.67 & 27.50 & 25.08 \\ 
  City of London & 32.84 & 33.35 & 30.40 & 35.06 & 36.41 & 38.71 & 32.44 & 37.14 & 34.57 & 35.67 & 32.27 & 37.26 & 37.33 \\ 
  Kensington \& Chelsea & 25.99 & 29.27 & 24.24 & 29.67 & 27.69 & 32.60 & 25.34 & 31.47 & 26.60 & 30.92 & 24.91 & 30.02 & 29.51 \\ 
  Southwark & 21.92 & 24.81 & 20.07 & 27.51 & 23.99 & 30.99 & 21.48 & 28.13 & 22.32 & 23.59 & 20.60 & 27.98 & 24.78 \\ 
  Barking \& Dagenham & 30.37 & 30.42 & 29.03 & 39.18 & 33.11 & 33.17 & 31.47 & 40.79 & 30.77 & 31.79 & 29.28 & 40.19 & 37.73 \\ 
  Barnet & 21.06 & 23.16 & 18.60 & 27.63 & 22.92 & 30.49 & 19.73 & 28.85 & 21.91 & 23.76 & 19.55 & 28.61 & 22.70 \\ 
  Brent & 26.56 & 27.69 & 25.39 & 31.04 & 28.37 & 31.83 & 26.82 & 37.83 & 26.07 & 27.44 & 24.72 & 32.19 & 31.54 \\ 
  Ealing & 25.01 & 26.85 & 23.18 & 31.91 & 26.42 & 29.32 & 24.04 & 33.36 & 24.71 & 27.78 & 23.18 & 33.50 & 27.78 \\ 
  Greenwich & 22.07 & 25.82 & 19.94 & 28.06 & 23.67 & 33.58 & 20.78 & 30.29 & 22.94 & 25.33 & 20.83 & 28.56 & 23.88 \\ 
  Hackney & 27.85 & 29.39 & 26.36 & 29.88 & 29.84 & 29.82 & 27.48 & 32.30 & 27.78 & 28.29 & 26.30 & 30.79 & 31.88 \\ 
  Haringey & 21.66 & 25.00 & 19.74 & 29.73 & 23.47 & 26.88 & 20.91 & 33.23 & 22.55 & 23.15 & 19.98 & 31.28 & 24.20 \\ 
  Havering & 24.72 & 27.85 & 22.13 & 30.80 & 25.90 & 30.99 & 23.23 & 32.98 & 26.82 & 27.41 & 24.07 & 31.00 & 26.62 \\ 
  Hillingdon & 26.27 & 27.76 & 23.68 & 31.74 & 28.47 & 31.93 & 25.37 & 34.14 & 27.08 & 27.28 & 25.01 & 31.88 & 29.41 \\ 
  Kingston upon Thames & 26.78 & 28.53 & 24.17 & 30.15 & 29.01 & 31.79 & 25.22 & 32.63 & 27.92 & 32.06 & 25.75 & 30.94 & 29.08 \\ 
  Merton & 26.34 & 28.72 & 24.61 & 29.95 & 28.88 & 37.73 & 26.21 & 33.45 & 27.66 & 28.55 & 25.31 & 31.09 & 30.39 \\ 
  Newham & 18.97 & 20.27 & 17.28 & 24.50 & 20.72 & 25.60 & 18.72 & 25.80 & 21.39 & 22.63 & 19.93 & 26.62 & 21.31 \\ 
  Redbridge & 23.25 & 24.66 & 21.00 & 28.19 & 24.66 & 26.55 & 21.94 & 32.35 & 23.83 & 24.95 & 21.78 & 28.66 & 24.52 \\ 
  Lewisham & 22.13 & 25.41 & 20.05 & 26.08 & 23.88 & 30.23 & 21.28 & 27.71 & 23.51 & 23.93 & 21.68 & 27.19 & 24.19 \\ 
  Richmond upon Thames & 33.10 & 36.05 & 31.35 & 36.11 & 35.30 & 37.29 & 33.35 & 39.70 & 33.68 & 34.51 & 31.44 & 36.92 & 39.78 \\ 
  Hounslow & 24.66 & 26.01 & 22.83 & 34.07 & 27.10 & 28.63 & 24.36 & 35.88 & 29.58 & 34.09 & 27.45 & 38.50 & 28.78 \\ 
  Croydon & 21.89 & 24.95 & 19.59 & 29.90 & 23.46 & 27.67 & 20.32 & 31.37 & 23.54 & 25.66 & 20.78 & 29.03 & 23.29 \\ 
  Enfield & 22.89 & 25.90 & 20.40 & 30.46 & 24.45 & 27.57 & 21.64 & 31.05 & 24.49 & 26.03 & 21.57 & 31.65 & 24.90 \\ 
  WalthamForest & 22.74 & 24.91 & 20.59 & 28.85 & 24.86 & 27.77 & 21.76 & 31.67 & 22.79 & 26.95 & 20.92 & 29.07 & 25.31 \\ 
  Harrow & 26.05 & 26.42 & 23.96 & 30.65 & 28.10 & 33.57 & 25.19 & 32.95 & 27.02 & 30.99 & 24.74 & 31.68 & 29.93 \\ 
  Bromley & 27.41 & 29.33 & 25.37 & 33.56 & 29.11 & 29.45 & 26.99 & 36.16 & 29.51 & 30.75 & 27.24 & 33.56 & 31.20 \\ 
  Sutton & 33.36 & 35.88 & 31.06 & 38.71 & 35.93 & 39.90 & 32.82 & 41.06 & 34.09 & 34.96 & 31.92 & 37.64 & 39.06 \\ 
  Bexley & 26.59 & 28.40 & 24.31 & 35.34 & 28.19 & 33.72 & 25.47 & 38.06 & 27.18 & 29.17 & 25.47 & 33.17 & 29.08 \\[0.2cm] 
  Average & 24.96 & 27.14 & 22.96 & 30.48 & 26.98 & 31.03 & 24.36 & 32.80 & 26.09 & 28.00 & 24.08 & 31.46 & 28.28 \\ 
  \hline
\end{tabular}
\end{minipage} }
\end{table}

Table \ref{tab:SMAPE_all_boroughs} reports SMAPEs for all boroughs across the  updating procedures and the  ML methods.
Let us first investigate the monitoring procedures, namely ours versus the changepoint test, across the ML methods.
Our monitoring approach compares favorably, across all boroughs, against changepoint-based monitoring. For example, the random forest average SMAPE (last row  in Table \ref{tab:SMAPE_all_boroughs}) of 24.96\% increases to 27.14\% for Pelt. The difference is statistically significant (paired $t$-tests at 5\% throughout this section) for random forest, XGBoost and lasso. Since both monitoring schemes are applied to the same ML algorithms, these discrepancies are caused by the fact that Pelt detects less instabilities than the standard equality of means test: Pelt detects only between 8 and 21 breaks (across the boroughs and ML algorithms), whereas ours detects between 74 and 153 breaks. 
Moreover, any changepoint-based monitoring procedure typically faces a detection delay between the date on which a break is \textit{detected} and the date the break  \textit{occurred}, which is around 7 to 13 days for Pelt (on average). Our monitoring procedure, in contrast, does not face such a detection delay. Further details on the monitoring procedure results are provided in Section \ref{subsec:results:monitor}.

Zooming into the performance of our monitoring procedure across the ML algorithms, we notice that random forest with an average SMAPE of 24.96\% outperforms with a few percentage points XGBoost and lasso (respectively 26.98\% and 26.09\%). These differences are statistically significant. 
It is remarkable that for each individual borough, monitoring with random forest mildly yet consistently dominates the other ML algorithms.

Next, we focus on the performance compared to the other benchmarks. The monitored ML algorithms clearly outperform the Naive approach. For example, random forest has an average SMAPE that is 12 percent lower than this approach. Also deterministic semester re-training results in systematically higher SMAPEs for all boroughs and ML algorithms. For example, random forest monitoring has an average SMAPE that is 18 percent lower than deterministic re-training. These results are statistically significant.
Finally, as expected, the in practice computationally unfeasible setting of daily re-training yields the lowest SMAPEs for all ML algorithms; the difference with our monitoring procedure are statistically significant. For example, daily random forest has an average SMAPE of 22.96\%, two percentage points lower than our monitoring procedure. These differences also systematically occur for individual boroughs. 

Although monitoring-based re-training pays a price in terms of forecast accuracy compared to daily re-training, the former does allow one to save the random forest daily 5-minute training time (on average per borough) during periods of forecast stability where  forecasts are produced with a constant forecast function (and the equality of means tests is computed instantaneously).\footnote{We apply off-the-shelf ML algorithms with default tuning parameters as described in Section \ref{subsec:treeML} since hyper-parameter tuning is not the main objective of this paper. If one additionally tunes hyper-parameters in practice, computing time considerably increases further.} 
While at first sight this might seem a modest saving for the sample of 32 delivery boroughs we consider, such savings quickly and considerably add up when implemented for the entire market of the digital platform, which easily consists of hundreds to thousands (considering expansion) of delivery areas. Our monitoring procedure offers a flexible, automated way to smear out computing time over days since areas rarely require simultaneous re-training (see Section \ref{subsec:results:monitor}) due to the heterogeneous market composition. 

Finally, we return to Figure \ref{fig:heat_maps_london} (bottom panel) which displays the SMAPE London heatmap for the random forest monitoring procedure. SMAPEs range  between 18\% (Tower Hamlets) and 33\% (Sutton). There is no specific dependence between geographic location and forecast performance. However, comparing with Figure \ref{fig:heat_maps_london} (top panel), we identify a strong negative relationship between demand volume and SMAPE, with a linear correlation coefficient of -0.93. This is important information when the platform business optimizes the size of its delivery zones, keeping into account that too large zones are inefficient for managing and compensating couriers, i.e.\ the supply side of the platform.

\subsection{Monitoring Procedure Insights} \label{subsec:results:monitor}
The  monitoring procedure allows computationally efficient re-training of ML algorithms at dates where forecast loss streams indicate instability. In contrast to re-training at a pre-specified frequency, these dates are unknown a priori. We now report details on the frequency and timing of ML algorithm re-training.  

\begin{figure}[t]
\centering
\includegraphics[width=0.85\textwidth]{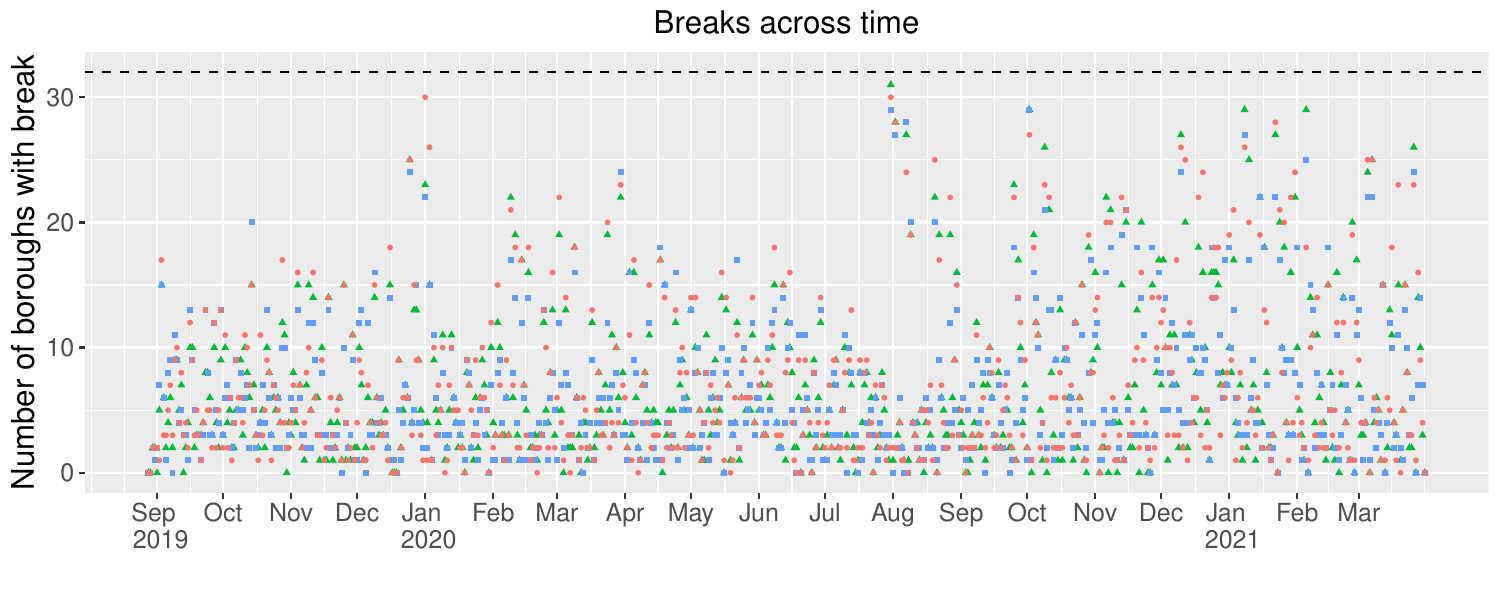}
\includegraphics[width=0.85\textwidth]{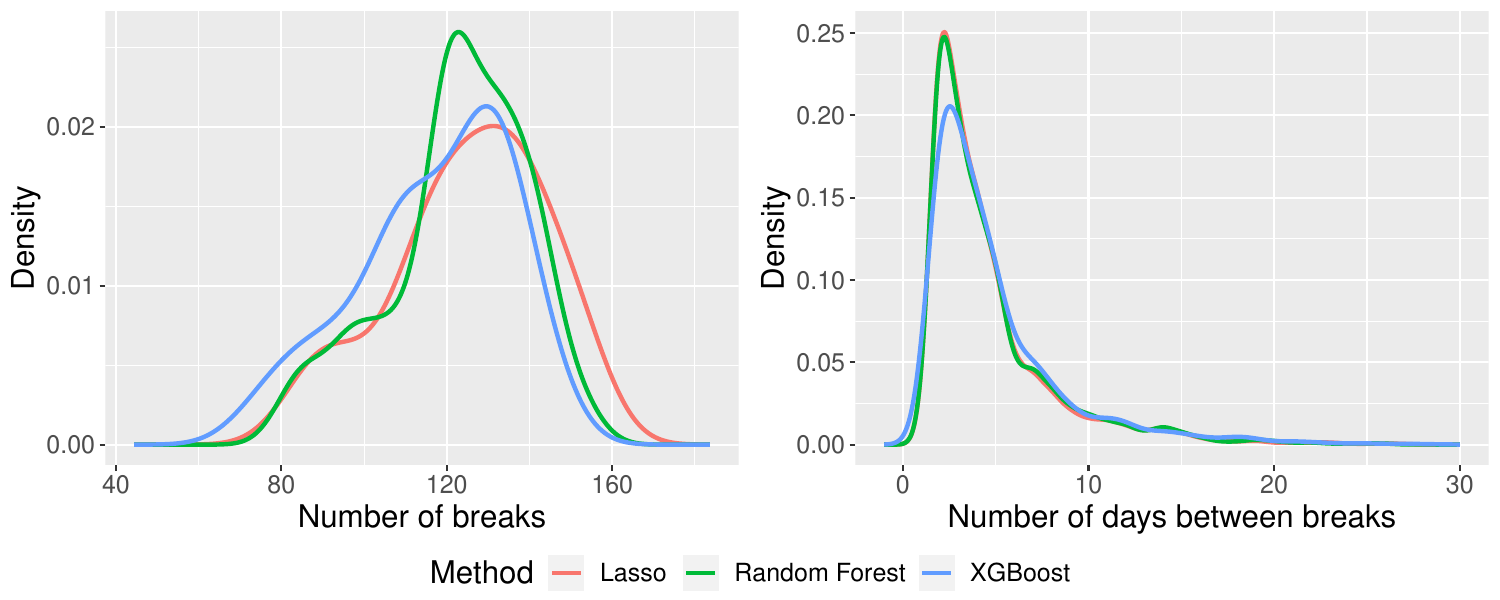}
\caption{Number of breaks for the random forest (green), XGBoost (blue) and lasso (red) detected across time (top) together with density estimates for the number of breaks and number of days between breaks (bottom). 
\label{fig:break_information}}
\end{figure}

Figure \ref{fig:break_information} (top panel) displays for each day in the out-of-sample evaluation period the number of boroughs where the monitoring test is rejected for random forest (green), XGBoost (blue) and lasso (red). 
The horizontal upper bound at 32 implies daily re-training for each borough. Overall, the number of rejected monitoring tests (in short referred to as ``breaks" in the remainder) are rarely close to this upper bound. The number of breaks per day is relatively concentrated and often below 10 until August 2020 after which it rises regularly above 20. The color scheme 
shows that the monitoring tests behave similarly across ML algorithms in terms of number of breaks, although some days show large discrepancies.

Next, we summarize the borough-specific break results. Figure \ref{fig:break_information} (bottom panel) shows that the number of breaks per borough for each ML algorithm is centered around 120 and is highly variable among the boroughs. The duration, i.e.\ the number of days between breaks, shares the same right skewed pattern across ML algorithms. Although the ML algorithms require frequent re-training there are also boroughs with segments of stability mounting to 30 days, thereby calling for a borough-specific monitoring procedure. 

Since our monitoring procedure relies on a simple equality of means test, a natural question to ask is how sensitive the results are to the size of the test.
We therefore re-did the analysis with a significance level of 1\% instead of 5\%. The results are summarized in Table \ref{tab:1percentmonitor} of  Appendix \ref{app:results}. 
The test rejects less often for all ML algorithms; for example, the number of breaks in the  random forest forecast loss stream  goes down  from  123 to 72, on average, over the boroughs. This substantial gain in computing time per borough, i.e. less frequent retraining, comes at a small cost of higher forecast errors. For instance, on average over the boroughs, SMAPE performance increases from 24.96\% to 25.34\% for random forest.

\subsection{Spillovers Across Boroughs and Variable Importance} \label{subsec:results:shap}
\begin{figure}[t]
\centering
\includegraphics[width=0.75\textwidth]{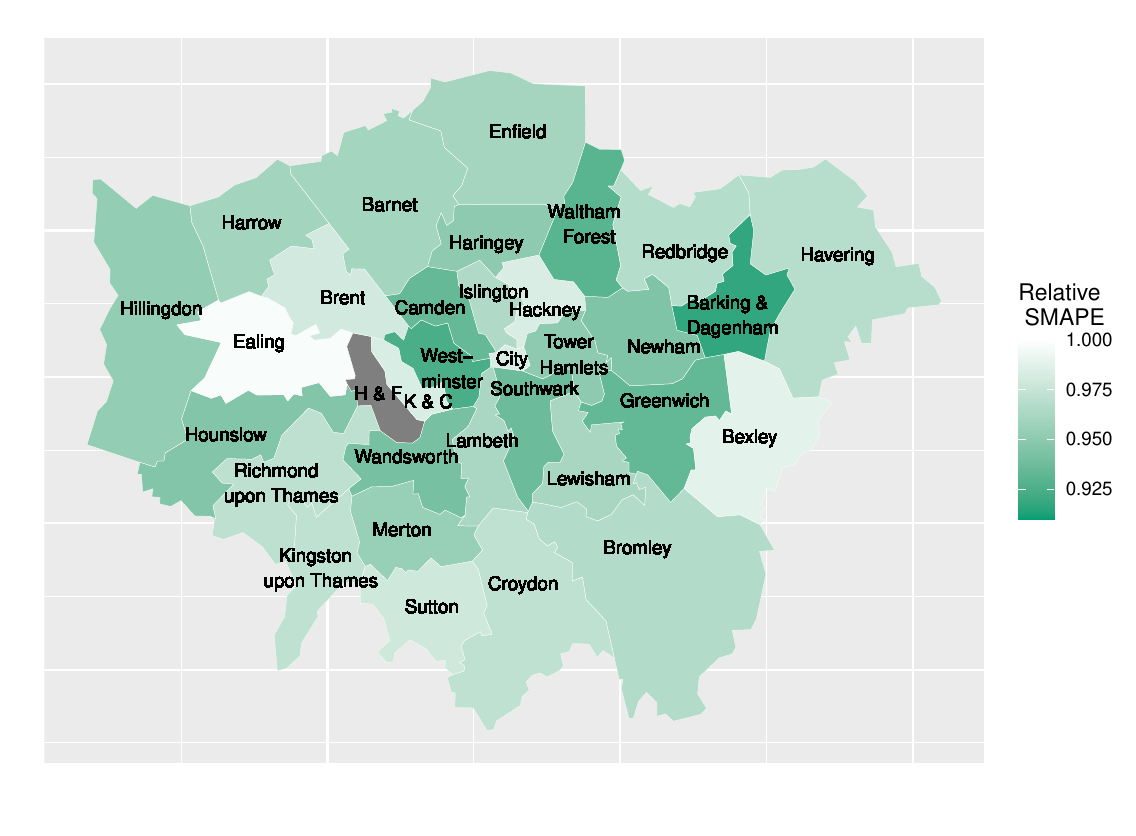}
\vspace{-1cm}
\caption{Relative SMAPE performance heat map for random forest monitoring with versus without boroughs included in the information set.} \label{fig:SMAPE_monitor_RF_others_vs_own}
\end{figure}

To gauge whether other boroughs include useful information when forecasting demand in a specific focal borough, we consider a ``univariate" model where each ML algorithm is trained with \textit{own} borough-specific past data only, trend and seasonality. Figure \ref{fig:SMAPE_monitor_RF_others_vs_own} 
displays a heatmap of SMAPE relative performance of random forest monitoring  where green (white) indicates better (equal) forecast performance when past information of other boroughs is included versus the inclusion of own borough-specific data only. 
We focus on random forest as it is the best performing ML algorithm, but the findings also hold for the other ML algorithms.
The heat map clearly indicates that the inclusion of other borough information implies lower, hence better, SMAPEs, except for  Ealing where the difference is negligible. 
The reduction in SMAPE by adding other boroughs in the information set is statistically significant for each of the three ML algorithms.
The largest SMAPE difference in favor of including other borough lags amounts to 2.73 percentage points for Barking \& Dagenham.

\begin{figure}[t]
\centering
\includegraphics[width=0.8\textwidth]{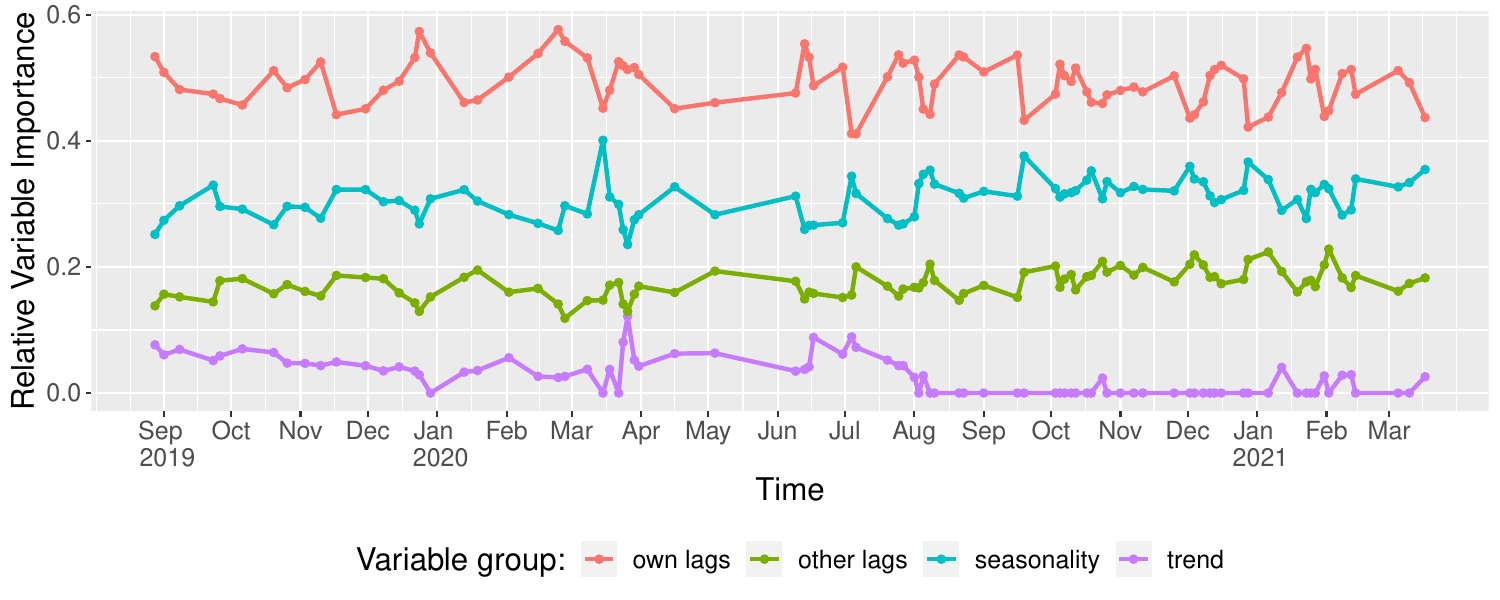}
\includegraphics[width=0.8\textwidth]{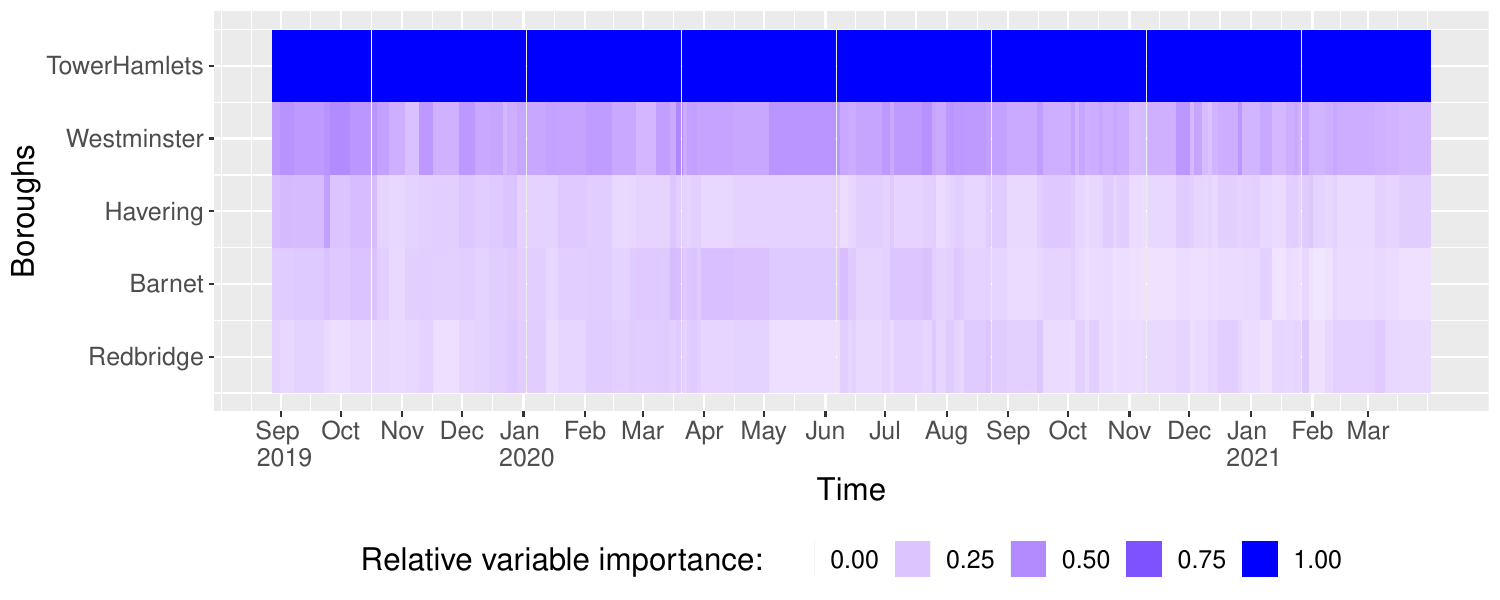}

\caption{City of London: Relative variable importance across time and by variable group (top) or by the five most important other boroughs (bottom) for random forest monitoring.} \label{fig:shap_values_vi}
\end{figure}

We next aim to better understand how the different groups-- own lags, other lags, seasonality and trend --of the information set drive the forecasting power of the random forest monitoring model.  
To this end, we compute SHAP values \citep{shapley1953value, lundberg2017unified},  considered to be the current state-of-the-art method for interpreting ML models, via \texttt{treeshap} \citep{treeshapR} in \texttt{R}, a fast implementation for tree ensemble models \citep{lundberg2018consistent}. We refer the interested reader to \cite{molnar2020interpretable} for a detailed introduction to SHAP values.
For each time point in the out-of-sample evaluation period where the random forest monitoring procedure detects a break, and hence updates the forecast function, we compute the SHAP values for all observations in the 15-min day-ahead batch.

Figure \ref{fig:shap_values_vi} (top panel) displays relative variable importance for City of London across time and by variable group. 
City of London own historical demand data contributes most to the forecast performance, namely about 50\%, followed by seasonality  with percentages around 30\%. Both groups hover around these levels with 
some fluctuation over time but consistently remain the dominant generators of predictability. 
Trend effects count initially for about 10\% but gradually decline until the fall of 2020 after which they become an irrelevant driver for forecast performance. 
Finally, historical information from other boroughs weights close to 20\% on average, and slightly increases over time in line with the higher average correlations late 2020 in Figure \ref{fig:tv_correlations}. 
In this regard, Figure \ref{fig:shap_values_vi} (bottom panel) shows the five most important boroughs for City of London, and highlights the 
(relatively time invariant) main importance of the neighboring Tower Hamlets and Westminster boroughs. The contribution of the Havering, Barnet and Redbridge, however, varies mildly over time.

\section{Conclusion} \label{Conclusion}
Supervised machine learning algorithms such as random forest, XGboost and lasso, among others, are becoming intensively used by data analysts in prediction/forecasting problems. The main reasons are their documented successes and the fact that these algorithms are readily available in software packages with minimal implementation efforts. However, 
once deployed in high-frequency large-scale  streaming data settings-- which frequently arise nowadays across a wide range of application areas --an ML algorithm becomes computationally unfeasible to re-train each time new data batches come in. 
To prevent forecast performance deterioration, we propose a monitoring procedure that signals when re-training of the algorithm is required because of a detected instability in the streaming forecast loss function. The process amounts to testing if the forecast loss of a new incoming data batch significantly differs from a reference loss batch. The implementation is easy and applicable to forecast functions coming from one's favorite ML algorithm.

On-demand platforms recently generated considerable research interest in business, economics and management (e.g., \citealp{Burtch_mansci2018, helfat2018dynamic, Chen_JPE_2019}), and they naturally face high-frequency large-scale data stream settings subject to frequent changes making them an appealing case study to demonstrate the value of our monitoring procedure. 
We use a dataset consisting of 15-minute frequency demand streams in 32 boroughs of London. 
Each borough forecast function is trained using its own and others borough's historical information. In the evaluation period from  August 2019 to March 2021, the number of times the algorithm needs re-training roughly varies between 50 and 160 across boroughs. The results show that the timing of re-training is aligned across the three ML algorithms.  In terms of forecast performance, random forest monitoring dominates.

In this paper, we test if an incoming forecast loss batch is different from a reference batch. To have a decent incoming batch size, we test and re-train (if needed)  end-of-day, but not within a business day. It would be an interesting future research extension to investigate if intra-day monitoring (possibly based on more fine-grained data than 15-min slots) further enhances forecast performance. 
Besides, while we demonstrate the proposed monitoring procedure with standard, popular ML algorithms,  
another extension is to investigate if forecast performance could be further improved by using, for instance, dedicated random forest techniques for spatially dependent data as in \cite{saha2021random}. In fact, since it turns out from our off-the-shelf random forest variable importance study that geographically close boroughs are more important than remote boroughs, explicitly accounting for this dependence can potentially further improve the precision of the forecast streams.

Finally, though we have focused on demand forecasting at a logistics platform, it is important to note that our monitoring procedure is general and has further potential in  a wide range of other application fields 
requiring monitoring high-frequency  forecast loss streams and detecting instability at large scale,  in a computationally feasible manner.
As part of their recommender systems to better understand consumer behavior and purchasing behavior,  e-commerce businesses and online market places such as Amazon or Walmart require fast forecasts  of
bursty streaming web traffic data at high-frequency intervals that can rapidly adapt to structural changes. 
Other data applications include, amongst others,
forecasting intra-day residential electricity consumption data which have become available due to the increasing adoption of smart meters, 
forecasting tick-by-tick financial data which attracts growing attention from governmental regulators and industry, 
or
forecasting streaming air-pollution data on a fine-resolution temporal and spatial scale to provide insights into air quality at local environments.

\begingroup
\setstretch{0.05}
\linespread{0.5}
\bibliographystyle{apalike}
\bibliography{RomWil}
\endgroup		
\clearpage

\newpage

\renewcommand{\thetable}{A.\arabic{table}}
\setcounter{table}{0}

\renewcommand{\thefigure}{A.\arabic{figure}}
\setcounter{figure}{0}

\begin{appendices}

\section{Platform Application: Additional Results} \label{app:results}

\begin{table}[ht]
\caption{Monitoring results (SMAPE and number of detected breaks) with a significance level of the standard equality of means test of 5\% vs 1\%. 
\label{tab:1percentmonitor}}
\resizebox{0.76\textwidth}{!}{\begin{minipage}{\textwidth}
\begin{tabular}{lcccccccccccc}
  \hline
Borough  & \multicolumn{6}{c}{\underline{SMAPE}} & \multicolumn{6}{c}{\underline{Number of Breaks}}\\
  & \multicolumn{2}{c}{RF}  & \multicolumn{2}{c}{XGBoost} & \multicolumn{2}{c}{Lasso} & \multicolumn{2}{c}{RF}  & \multicolumn{2}{c}{XGBoost} & \multicolumn{2}{c}{Lasso} \\
 &  5\% &  1\% &  5\% &  1\% &  5\% &  1\%  &  5\% &  1\% 
 & 5\% &  1\%  & 5\% &  1\%  \\ 
  \hline
TowerHamlets & 18.07 & 18.12 & 20.02 & 20.79 & 19.47 & 19.92 & 95.00 & 59.00 & 98.00 & 57.00 & 93.00 & 58.00 \\ 
  Wandsworth & 21.38 & 21.57 & 23.51 & 25.14 & 22.60 & 22.96 & 131.00 & 80.00 & 135.00 & 77.00 & 140.00 & 84.00 \\ 
  Camden & 22.25 & 22.59 & 24.90 & 25.32 & 23.46 & 23.65 & 121.00 & 75.00 & 111.00 & 60.00 & 112.00 & 69.00 \\ 
  Islington & 31.55 & 31.68 & 34.27 & 34.57 & 32.88 & 33.40 & 124.00 & 71.00 & 102.00 & 63.00 & 122.00 & 74.00 \\ 
  Westminster & 20.87 & 21.87 & 22.72 & 23.73 & 22.53 & 23.93 & 97.00 & 56.00 & 84.00 & 50.00 & 95.00 & 58.00 \\ 
  Lambeth & 22.01 & 21.92 & 24.02 & 24.83 & 23.63 & 23.97 & 144.00 & 84.00 & 136.00 & 85.00 & 151.00 & 95.00 \\ 
  City of London & 32.84 & 32.85 & 36.41 & 35.87 & 34.57 & 35.20 & 87.00 & 52.00 & 82.00 & 52.00 & 87.00 & 53.00 \\ 
  Kensington \& Chelsea & 25.99 & 26.10 & 27.69 & 28.52 & 26.60 & 27.23 & 101.00 & 57.00 & 91.00 & 53.00 & 104.00 & 53.00 \\ 
  Southwark & 21.92 & 22.25 & 23.99 & 25.00 & 22.32 & 22.76 & 139.00 & 99.00 & 133.00 & 77.00 & 141.00 & 105.00 \\ 
B \& D & 30.37 & 32.15 & 33.11 & 33.41 & 30.77 & 31.87 & 83.00 & 36.00 & 74.00 & 36.00 & 87.00 & 46.00 \\ 
  Barnet & 21.06 & 23.29 & 22.92 & 24.76 & 21.91 & 25.05 & 130.00 & 79.00 & 129.00 & 70.00 & 134.00 & 67.00 \\ 
  Brent & 26.56 & 27.14 & 28.37 & 31.06 & 26.07 & 26.62 & 109.00 & 50.00 & 93.00 & 50.00 & 113.00 & 63.00 \\ 
  Ealing & 25.01 & 25.14 & 26.42 & 27.31 & 24.71 & 25.48 & 121.00 & 73.00 & 108.00 & 61.00 & 123.00 & 70.00 \\ 
  Greenwich & 22.07 & 22.73 & 23.67 & 24.54 & 22.94 & 22.72 & 144.00 & 85.00 & 144.00 & 80.00 & 146.00 & 87.00 \\ 
  Hackney & 27.85 & 28.14 & 29.84 & 29.61 & 27.78 & 27.92 & 121.00 & 71.00 & 123.00 & 69.00 & 126.00 & 72.00 \\ 
  Haringey & 21.66 & 21.82 & 23.47 & 26.06 & 22.55 & 22.83 & 134.00 & 86.00 & 130.00 & 74.00 & 130.00 & 84.00 \\ 
  Havering & 24.72 & 24.77 & 25.90 & 27.80 & 26.82 & 26.66 & 133.00 & 84.00 & 132.00 & 83.00 & 138.00 & 83.00 \\ 
  Hillingdon & 26.27 & 26.28 & 28.47 & 28.76 & 27.08 & 28.02 & 141.00 & 78.00 & 137.00 & 72.00 & 137.00 & 82.00 \\ 
  Kingston upon Thames & 26.78 & 27.24 & 29.01 & 31.20 & 27.92 & 28.90 & 120.00 & 68.00 & 114.00 & 63.00 & 129.00 & 68.00 \\ 
  Merton & 26.34 & 26.19 & 28.88 & 30.14 & 27.66 & 27.44 & 120.00 & 65.00 & 109.00 & 53.00 & 119.00 & 69.00 \\ 
  Newham & 18.97 & 19.59 & 20.72 & 21.73 & 21.39 & 21.67 & 119.00 & 69.00 & 116.00 & 67.00 & 123.00 & 81.00 \\ 
  Redbridge & 23.25 & 23.24 & 24.66 & 25.49 & 23.83 & 23.88 & 138.00 & 91.00 & 144.00 & 91.00 & 137.00 & 91.00 \\ 
  Lewisham & 22.13 & 22.29 & 23.88 & 23.76 & 23.51 & 23.77 & 152.00 & 84.00 & 128.00 & 83.00 & 153.00 & 90.00 \\ 
  Richmond upon Thames & 33.10 & 33.18 & 35.30 & 35.78 & 33.68 & 33.84 & 106.00 & 54.00 & 106.00 & 49.00 & 111.00 & 58.00 \\ 
  Hounslow & 24.66 & 25.70 & 27.10 & 30.34 & 29.58 & 29.90 & 121.00 & 61.00 & 108.00 & 54.00 & 119.00 & 66.00 \\ 
  Croydon & 21.89 & 21.69 & 23.46 & 24.02 & 23.54 & 23.03 & 128.00 & 82.00 & 127.00 & 72.00 & 132.00 & 76.00 \\ 
  Enfield & 22.89 & 23.42 & 24.45 & 25.29 & 24.49 & 24.76 & 134.00 & 87.00 & 130.00 & 87.00 & 138.00 & 89.00 \\ 
  WalthamForest & 22.74 & 22.98 & 24.86 & 25.35 & 22.79 & 23.28 & 125.00 & 81.00 & 113.00 & 82.00 & 133.00 & 86.00 \\ 
  Harrow & 26.05 & 26.55 & 28.10 & 28.98 & 27.02 & 27.35 & 113.00 & 63.00 & 119.00 & 60.00 & 120.00 & 62.00 \\ 
  Bromley & 27.41 & 28.02 & 29.11 & 30.61 & 29.51 & 30.04 & 119.00 & 70.00 & 127.00 & 55.00 & 113.00 & 64.00 \\ 
  Sutton & 33.36 & 33.78 & 35.93 & 36.32 & 34.09 & 34.99 & 131.00 & 72.00 & 123.00 & 72.00 & 150.00 & 80.00 \\ 
  Bexley & 26.59 & 26.60 & 28.19 & 29.24 & 27.18 & 27.38 & 141.00 & 86.00 & 137.00 & 75.00 & 151.00 & 86.00 \\ 
  &&&&&\\
  Average & 24.96 & 25.34 & 26.98 & 27.98 & 26.09 & 26.57 & 122.56 & 72.12 & 116.97 & 66.62 & 125.22 & 74.03 \\ 
   \hline
\end{tabular}
\end{minipage} }
\end{table}

\begin{table}[ht]
\centering
\caption{SMAPE results for monitoring procedure based on offline changepoint test Pelt versus online changepoint test FOCuS (with fixed threshold set at 8).} \label{Table-Pelt-vs-Focus}
\begin{tabular}{lcccccccccccc}
  \hline
Borough &&& \multicolumn{2}{c}{RF} &&& \multicolumn{2}{c}{XGBoost} &&& \multicolumn{2}{c}{Lasso} \\  
 &&& Pelt & FOCuS &&& Pelt & FOCuS &&& Pelt & FOCuS \\ 
  \hline
TowerHamlets &&& 21.39 & 21.38 &&& 23.39 & 31.36 &&& 22.40 & 24.57 \\ 
  Wandsworth &&& 23.44 & 24.69 &&& 28.21 & 31.29 &&& 26.81 & 36.21 \\ 
  Camden &&& 25.61 & 23.83 &&& 30.24 & 29.79 &&& 25.07 & 25.58 \\ 
  Islington &&& 33.85 & 32.99 &&& 35.39 & 35.55 &&& 34.59 & 36.30 \\ 
  Westminster &&& 22.52 & 23.42 &&& 28.02 & 26.47 &&& 24.35 & 24.76 \\ 
  Lambeth &&& 24.75 & 24.49 &&& 29.89 & 26.67 &&& 25.31 & 27.71 \\ 
  City of London &&& 33.35 & 38.95 &&& 38.71 & 41.04 &&& 35.67 & 43.95 \\ 
  Kensington \& Chelsea &&& 29.27 & 29.26 &&& 32.60 & 35.65 &&& 30.92 & 30.37 \\ 
  Southwark &&& 24.81 & 25.02 &&& 30.99 & 32.69 &&& 23.59 & 24.81 \\ 
  Barking \& Dagenham &&& 30.42 & 32.12 &&& 33.17 & 35.89 &&& 31.79 & 32.70 \\ 
  Barnet &&& 23.16 & 23.38 &&& 30.49 & 31.27 &&& 23.76 & 31.59 \\ 
  Brent &&& 27.69 & 27.94 &&& 31.83 & 32.37 &&& 27.44 & 27.73 \\ 
  Ealing &&& 26.85 & 25.74 &&& 29.32 & 30.97 &&& 27.78 & 33.29 \\ 
  Greenwich &&& 25.82 & 27.46 &&& 33.58 & 35.97 &&& 25.33 & 33.31 \\ 
  Hackney &&& 29.39 & 30.29 &&& 29.82 & 35.29 &&& 28.29 & 29.27 \\ 
  Haringey &&& 25.00 & 24.07 &&& 26.88 & 31.53 &&& 23.15 & 28.78 \\ 
  Havering &&& 27.85 & 29.06 &&& 30.99 & 36.15 &&& 27.41 & 35.31 \\ 
  Hillingdon &&& 27.76 & 32.25 &&& 31.93 & 32.30 &&& 27.28 & 35.16 \\ 
  Kingston upon Thames &&& 28.53 & 29.31 &&& 31.79 & 35.33 &&& 32.06 & 30.06 \\ 
  Merton &&& 28.72 & 30.29 &&& 37.73 & 34.56 &&& 28.55 & 33.28 \\ 
  Newham &&& 20.27 & 23.48 &&& 25.60 & 29.83 &&& 22.63 & 31.27 \\ 
  Redbridge &&& 24.66 & 28.15 &&& 26.55 & 31.46 &&& 24.95 & 34.05 \\ 
  Lewisham &&& 25.41 & 28.30 &&& 30.23 & 28.51 &&& 23.93 & 26.43 \\ 
  Richmond upon Thames &&& 36.05 & 35.44 &&& 37.29 & 37.03 &&& 34.51 & 33.90 \\ 
  Hounslow &&& 26.01 & 26.06 &&& 28.63 & 30.07 &&& 34.09 & 36.45 \\ 
  Croydon &&& 24.95 & 25.01 &&& 27.67 & 31.58 &&& 25.66 & 27.01 \\ 
  Enfield &&& 25.90 & 26.06 &&& 27.57 & 33.05 &&& 26.03 & 38.23 \\ 
  WalthamForest &&& 24.91 & 24.85 &&& 27.77 & 29.92 &&& 26.95 & 30.81 \\ 
  Harrow &&& 26.42 & 28.49 &&& 33.57 & 32.34 &&& 30.99 & 30.35 \\ 
  Bromley &&& 29.33 & 29.73 &&& 29.45 & 30.47 &&& 30.75 & 30.30 \\ 
  Sutton &&& 35.88 & 37.04 &&& 39.90 & 42.41 &&& 34.96 & 37.56 \\ 
  Bexley &&& 28.40 & 29.19 &&& 33.72 & 39.08 &&& 29.17 & 32.75 \\ 
  \\
  Average &&& 27.14 & 28.05 &&& 31.03 & 33.06 &&& 28.00 & 31.68 \\ 

   \hline
\end{tabular}
\end{table}

\end{appendices}
\end{document}